\documentclass[letterpaper, 10pt]{article}
\usepackage[utf8]{inputenc}
\usepackage[top=1in, bottom=1in, left=1in, right=1in]{geometry}

\RequirePackage[explicit]{titlesec}
\titlespacing*{\section}{0pt}{0.5em}{0.3pt}
\titlespacing*{\subsection}{0pt}{0.35em}{0pt}
\titlespacing*{\subsubsection}{0pt}{0.25em}{0pt}

\usepackage{caption}
\RequirePackage{fancyhdr}
\usepackage{lastpage}

\usepackage{mathpazo}
\usepackage{xcolor}
\usepackage[
  colorlinks=true,
  linkcolor=blue,
  anchorcolor=blue,
  citecolor=blue,
  filecolor=blue,
  menucolor=blue,
  runcolor=blue,
  urlcolor=blue]{hyperref}
\usepackage{amsmath, amsthm, amssymb, amsfonts}
\usepackage{upgreek}
\usepackage{dsfont}

\usepackage{enumerate}
\usepackage{enumitem}

\usepackage{algorithm}
\usepackage{algpseudocode}

\setlist[enumerate]{topsep=1ex, itemsep=.75ex, partopsep=1ex, parsep=1ex}

\usepackage{graphicx}
\graphicspath{{figures/}}
\usepackage{subcaption}

\usepackage{multirow}

\usepackage[numbers, sort&compress]{natbib}

\title{Estimating Density, Velocity, and Pressure Fields in Supersonic Flows Using Physics-Informed BOS}

\author{
  Joseph P. Molnar$^1$,
  Lakshmi Venkatakrishnan$^2$,
  Bryan E. Schmidt$^3$,\\
  Timothy A. Sipkens$^4$, and
  Samuel J. Grauer$^{1,}$\thanks{Corresponding author: \href{mailto:sgrauer@psu.edu}{sgrauer@psu.edu}}\vspace*{.3em}\\
  {\small $^1$Department of Mechanical Engineering, Pennsylvania State University, University Park, PA}\vspace*{-.3em}\\
  {\small $^2$CSIR-National Aerospace Laboratories, Bangalore, India}\vspace*{-.3em}\\
  {\small $^3$Department of Mechanical and Aerospace Engineering, Case Western Reserve University, Cleveland, OH}\vspace*{-.3em}\\
  {\small $^4$Metrology Research Centre, National Research Council Canada, Ottawa, Canada}\vspace*{-1em}}
\date{}

\begin{document}
\maketitle
\setcounter{footnote}{2}

\begin{abstract}
We report a new workflow for background-oriented schlieren (BOS), termed ``physics-informed BOS,'' to extract density, velocity, and pressure fields from a pair of reference and distorted images. Our method uses a physics-informed neural network (PINN) to produce flow fields that simultaneously satisfy the measurement data and governing equations. For the high-speed, approximately inviscid flows of interest in this work, we specify a physics loss based on the Euler and irrotationality equations. BOS is a quantitative fluid visualization technique that is commonly used to characterize compressible flow. Images of a background pattern, positioned behind the measurement volume, are processed with computer vision and tomography algorithms to determine the density field. Crucially, BOS features a series of ill-posed inverse problems that require supplemental information (i.e., in addition to the images) to accurately reconstruct the flow. Current methods for BOS rely upon interpolation of the images or a penalty term to promote a globally- or piecewise-smooth solution. However, these algorithms are invariably incompatible with the flow physics, leading to errors in the density field. Physics-informed BOS directly reconstructs all the flow fields using a PINN that includes the BOS measurement model and governing equations. This procedure improves the accuracy of density estimates and also yields velocity and pressure data, which were not previously available. We demonstrate our approach by reconstructing synthetic data that corresponds to analytical and numerical phantoms as well as a single pair of experimental measurements. Our physics-informed reconstructions are significantly more accurate than conventional BOS estimates. Furthermore, to the best of our knowledge, this work represents the first use of a PINN to reconstruct a supersonic flow from experimental data of any kind.\par\vspace{.6em}

\end{abstract}
\vspace{.6em}

\section{Introduction}
\label{sec:intro}
Supersonic and hypersonic flows feature complex phenomena such as shock waves, shock wave--boundary layer interactions, and eddy shocklets, which must be considered in the design of next-generation aircraft and re-entry vehicles, projectiles, and combustion processes \cite{Dolvin2008}. Computational fluid dynamics (CFD) simulations play a vital role in engineering, but many vehicles are being designed to operate outside the parameter space in which engineering experience or CFD provide reliable means of analysis. Therefore, in order to support the design process, experimental measurements are needed to characterize and understand high-speed flow phenomena as well as to develop and validate numerical models.\par

Background-oriented schlieren (BOS) is a non-intrusive, quantitative flow visualization tool that can be applied to high-speed systems \cite{Raffel2015}. BOS has been widely used to characterize shock-laden flows \cite{Venkatakrishnan2004, Sommersel2008, Yamagishi2021, Gomez2022}, visualize combustion processes \cite{Grauer2018, Liu2022}, and estimate velocity fields \cite{Tokgoz2012}, amongst other applications \cite{Raffel2015}. The technique provides line-of-sight (LoS) integrated information about the flow via the apparent motion of a background pattern. Images of the pattern are distorted by refraction through the fluid, which is caused by density gradients along lines-of-sight from a background plate to the camera. Differences between a reference image, recorded before introducing the flow, and a distorted image from the experiment can thus be processed with a computer vision algorithm to render a ``synthetic schlieren'' image \cite{Dalziel2000}, which may reveal key fluid structures. Further, BOS data can be tomographically reconstructed to obtain a quantitative estimate of the density field. Unfortunately, BOS features a series of ill-posed inverse problems that admit an infinite set of solutions. Supplemental (or ``prior'') information is therefore needed to generate a unique, physical solution. Adding prior information to solve an inverse problem is termed ``regularization,'' and the aim of this work is to establish a physics-based approach to regularization in BOS.\par

\begin{figure}[ht]
    \centering
    \includegraphics[width=5.5in]{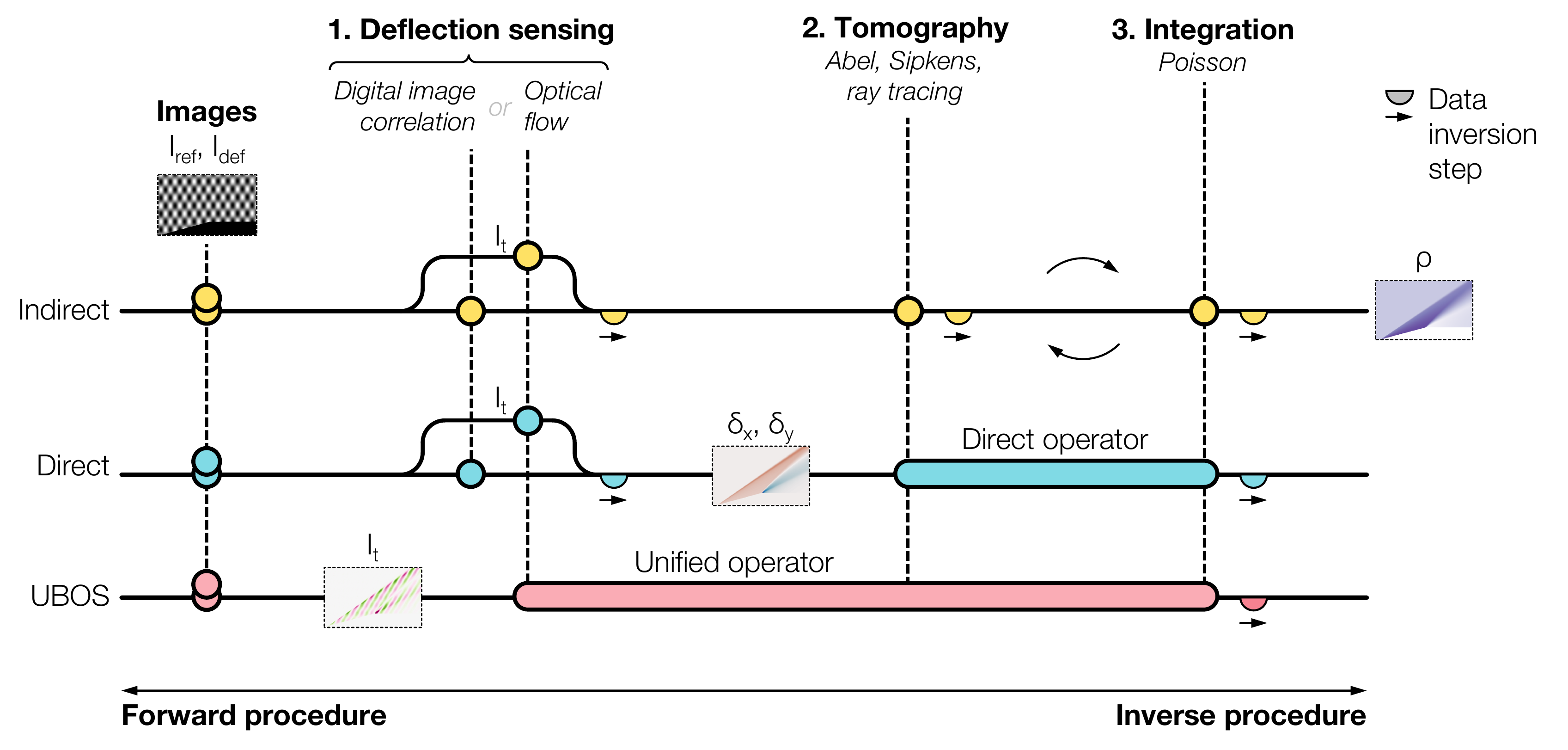}
    \caption{Graphical overview of quantitative BOS, in which image data is converted to a density field through deflection sensing, tomography, and a Poisson solver. The order of tomographic reconstruction and Poisson integration may be switched in indirect BOS or combined in direct BOS. In unified BOS (UBOS in the figure), all three procedures are performed by inverting a single operator.}
    \label{fig:BOS workflow}
\end{figure}

There are three inverse problems in BOS, depicted schematically in Fig.~\ref{fig:BOS workflow}, which are typically arranged in the following sequence. First, the image pair is converted to a set of deflection vectors through a procedure called deflection sensing. Here, a ``deflection'' is the 2D displacement of a point from the reference image to the same point in the distorted image. Deflections are usually resolved at the centroid of each pixel or interrogation window, although the deflection field is continuous, in principle. Second, individual components of the deflections are tomographically reconstructed, which yields gradients of the refractive index field. BOS is inherently sensitive to the refractive index field, but the density field is directly accessible for a fluid of constant composition via the Gladstone--Dale relation. Third, these reconstructions are incorporated into a Poisson equation that must be solved to recover the refractive index or density field, per se. Steps two and three can be reversed \cite{Rajendran2020} or combined \cite{Nicolas2016}, and it is possible to conduct all three steps at once via ``unified BOS'' \cite{Grauer2020}. Each step that is performed in isolation requires regularization, and it is advisable to combine steps, where possible, to reduce the reliance on prior information.\par

Deflection sensing is typically conducted using a cross-correlation \cite{Venkatakrishnan2004, Castner2012, Geerts2017}, dot tracking \cite{Rajendran2019}, or optical flow (OF) \cite{Atcheson2009, Heineck2020, Schmidt2021} algorithm. Cross-correlation methods identify the displacement of a multi-pixel window from the reference image to the distorted image. The use of multi-pixel windows, typically $8\times8$~px or larger, reduces the resolution of the deflection field \cite{Raffel1998, Atcheson2009, Schmidt2021}, which effectively amounts to a smoothing operation. Dot tracking algorithms attempt to determine the displacement of individual features on a pseudo-PIV background, which may be enhanced with a small-window, single-dot correlation step \cite{Rajendran2019}, but the resolution of this approach is limited by the density of dots, which is necessarily lower than the resolution of the sensor. Meanwhile, OF algorithms produce displacement fields with the same resolution as the original images and have been shown to outperform correlation and tracking algorithms in both resolution and accuracy \cite{Schmidt2021}. However, the OF problem features one equation and two unknowns per pixel, so an optimization criterion is needed to close the problem. Closure may be provided by assuming a locally-uniform solution (Lucas--Kanade OF \cite{Lucas1981}), global smoothness (Horn--Schunck OF \cite{Horn1981}), or a sparse representation in wavelet space \cite{Schmidt2021}. Lucas--Kanade OF is substantially similar to cross-correlation and does not yield much better estimates \cite{Atcheson2009}. The latter two techniques can generate accurate deflections, but it is exceedingly difficult to develop a physics-based constraint for OF in BOS due to the LoS-integrated nature of the deflection field \cite{Schmidt2021}.\par

Similar to OF, tomographic reconstruction inherently requires regularization to obtain a unique, physical solution and to counteract errors in the deflection estimates \cite{Daun2016}. Many examples of BOS feature an axisymmetric flow \cite{Raffel2015} such that the density field can be recovered from a single perspective via a modified Abel inversion \cite{Kogelschatz1972, Agrawal1999}. However, discrete analytical Abel inversion is unstable due to noise amplification that is inherent to numerical differentiation and a singularity at the line of symmetry \cite{Agrawal1999}. Better performance can be realized by coupling the forward model with an explicit penalty term and solving the resulting system with an optimization technique, often referred to as classical regularization \cite{Daun2006, Howard2016}. Far and away the most common penalties for tomography are the second-order Tikhonov \cite{Vauhkonen1998} and total variation (TV) \cite{Kolehmainen1998} terms, which promote global and piecewise smoothness, respectively. However, both penalties are exclusively minimized by a uniform field, which is incompatible with the density fields of interest in high-speed experimental fluid mechanics. This discrepancy introduces a trade-off between minimizing the measurement residuals and the penalty term. Consequently, both Tikhonov and TV regularization include a parameter that weights the penalty and must be carefully tuned for each experiment. While the use of a penalty term can stabilize tomographic reconstructions and improve their accuracy, errors associated with these penalty-based schemes are pervasive.\footnote{As an example, Tikhonov regularization is characterized by ``streaky'' artifacts, such as those in Fig.~5 of \cite{Wei2021}.} Regularized solutions tend to be overly-smooth, missing the fine detail present in a flow.\par

We propose a direct, physics-informed BOS workflow to avoid the regularization errors associated with deflection sensing and tomographic reconstruction and to recover the latent velocity and pressure fields. We utilize a physics-informed neural network (PINN) \cite{Raissi2019} to represent the flow, which requires data and physics loss terms. Our data loss is based on a unified BOS operator \cite{Grauer2020}, which directly relates a density field to raw image distortion data, and our physics loss comprises the compressible Euler and irrotationality equations. Previous work by the group of Karniadakis used a PINN to post-process 3D BOS tomography reconstructions of natural convection above an espresso cup \cite{Cai2021}. However, we found that directly embedding the measurement model into a PINN's data loss produces superior reconstructions \cite{Molnar2022}, leading to the present formulation.\par

This paper describes BOS, OF, tomography, and our physics-informed workflow. We apply the technique to reconstruct synthetic data from analytical and numerical phantoms as well as experimental images. Not only does physics-informed BOS yield better estimates of the density field than conventional techniques, it also generates estimates of the velocity and pressure fields. Furthermore, by processing noise-laden experimental LoS measurements with a realistic forward model instead of simulated point-wise data \cite{Mao2020, Jagtap2022}, this work represents an advance in the application of PINNs to high-speed flows.\par

\section{BOS}
\label{sec:BOS}
\begin{figure}[ht]
    \centering
    \includegraphics[width=4.25in]{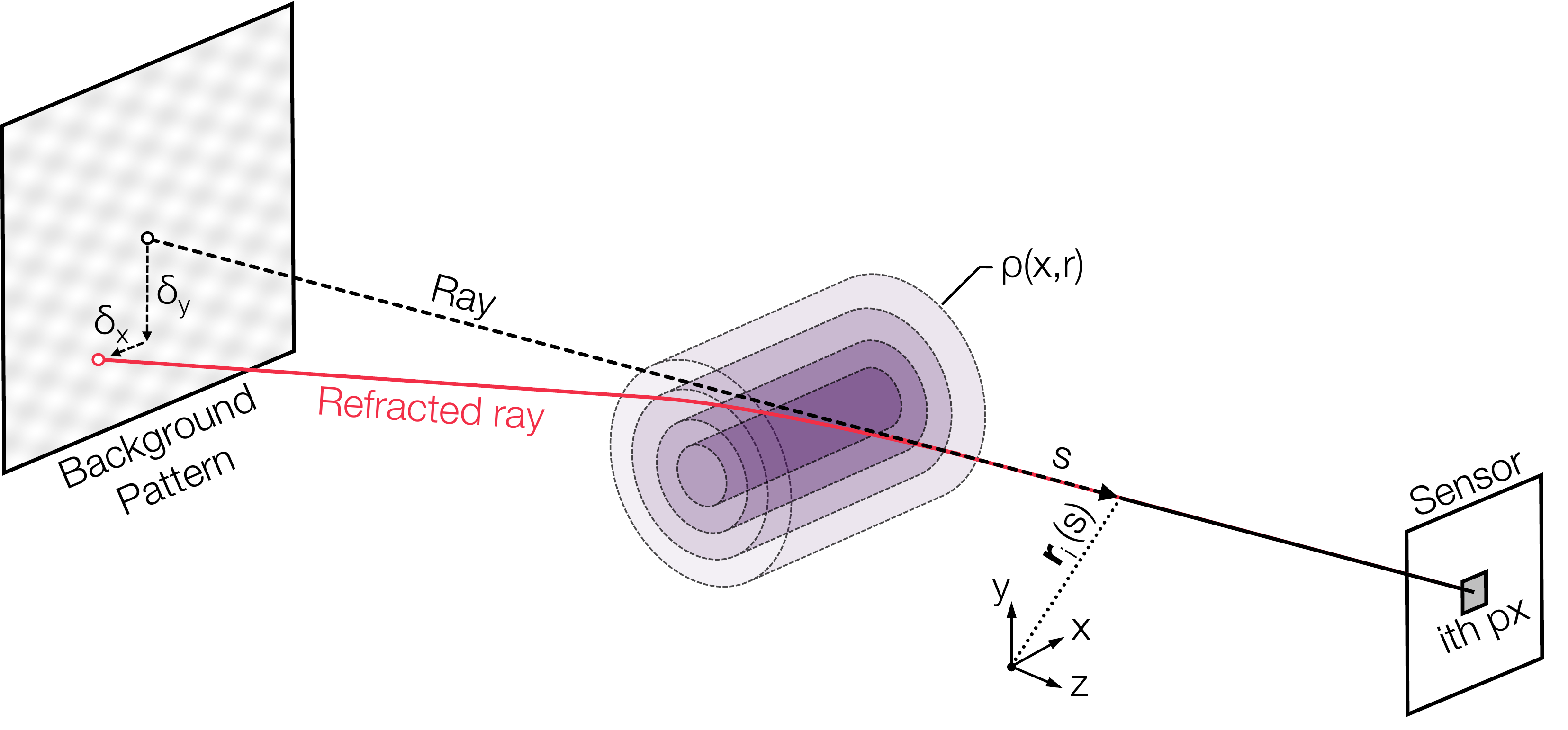}
    \caption{Schematic of a single-camera BOS setup for axisymmetric flows. Rays are refracted by density gradients in the flow, distorting images of the background pattern. Apparent deflections, $\boldsymbol\updelta = [\delta_\mathrm{x}, \delta_\mathrm{y}]^\mathrm{T}$, are determined with a computer vision algorithm and reconstructed to estimate $\rho$.}
    \label{fig:BOS setup}
\end{figure}

Figure~\ref{fig:BOS setup} depicts a common setup for axisymmetric BOS. A single camera is focused through the fluid to be measured onto a background plate that contains a pattern. Density gradients in the flow give rise to refractive index (speed of light) gradients, which cause wavefronts of light to bend (refract), thereby distorting images of the pattern. These distortions are characterized in terms of 2D displacement vectors, i.e., deflections, which are resolved at each pixel in the image. Deflections are estimated using a computer vision algorithm, and the deflection data may be reconstructed by inverting a BOS measurement model and/or solving a Poisson equation to estimate the unknown refractive index and density fields.\par

Throughout this work, we consider axisymmetric and planar flows that are aligned with a single background plate, as shown in Fig.~\ref{fig:BOS setup}. We assign the $x$-axis to the streamwise direction; the $x$- and $y$-axes mark the horizontal and vertical directions in the plane of the background plate; and the $z$-axis is normal to this plane. The camera is pointed directly towards the background and rotated such that the horizontal and vertical image coordinates coincide with the $x$- and $y$-axes, respectively, and the background pattern is assumed to be in focus. Continuous and discrete deflection models for BOS are introduced below, followed by an overview of deflection sensing with OF, the unified model used in this work, and axisymmetric reconstruction methods.\par

\subsection{Light propagation through variable index media}
\label{sec:BOS:continuous}
Wavefronts of light bend when the speed of light changes throughout a medium, which manifests as visible distortions of the background pattern in BOS. The speed of light in a medium is characterized by its refractive index, $n$. For gases, this property exhibits a linear dependence upon the material density, $\rho$, and composition, as described by the Gladstone--Dale equation,
\begin{equation}
    n = 1 + G\rho \Rightarrow \boldsymbol\nabla n = G\boldsymbol\nabla\rho.
    \label{equ:Gladstone-Dale}
\end{equation}
Here, $G$ is the Gladstone--Dale coefficient \cite{Gardiner1981}, which varies with chemical composition and exhibits a slight wavelength dependence. For measurements of air at visible wavelengths, $G \approx 2.26 \times 10^{-4}$~m$^3$/kg.\par

Light propagation is fundamentally governed by Maxwell's equations \cite{Born2013}. For a locally homogeneous region free of current and charge, these equations simplify to a wave equation. Assuming that variations in the refractive index field occur over much longer length scales than the wavelength of light, this wave equation simplifies to an eikonal equation, which describes the phase of light waves as a function of $n$, alone.\footnote{See Born and Wolf \cite{Born2013} for the derivation of an eikonal equation from Maxwell's equations.} In this limit, known as geometric optics, the propagation of light can be approximated by ``rays'' that travel normal to phase fronts of the wave. This phenomena is described by the so-called ray equation \cite{Stam1996},
\begin{equation}
    \frac{\mathrm{d}}{\mathrm{d}s} \left(n\frac{\mathrm{d}\mathbf{x}}{\mathrm{d}s}\right) = \boldsymbol\nabla n,
    \label{equ:ray equation}
\end{equation}
where $\mathbf{x}$ is the position of a massless particle traversing a ray of light and $s$ is a scalar progress variable, which indicates a distance along the ray as illustrated in Fig.~\ref{fig:BOS setup}.\par

The ray equation can be separated into two ordinary differential equations and integrated along a ray to calculate the deflection of that ray in the background plane, $\boldsymbol\updelta = [\delta_\mathrm{x}, \delta_\mathrm{y}]^\mathrm{T}$ \cite{Atcheson2008}. This integration follows a curved path, in principle, but the curve is slight within a BOS measurement volume \cite{Goldhahn2007}. Therefore, the paraxial assumption is invoked and the integral is carried out along the (straight) reference trajectory. For the camera setup shown in Fig.~\ref{fig:BOS setup}, the $\alpha$-direction deflection, for $\alpha \in \{x, y\}$, in pixel units is given by a path integral along the reference ray,
\begin{equation}
    \delta_\alpha = \frac{d \,\psi}{n_0} \int_\mathrm{ray} \nabla_\alpha \,n\mathopen{}\left[\mathbf{r}\mathopen{}\left(s\right)\right] \mathrm{d}s = \underbrace{\frac{d\,\psi\,G}{n_0}}_{C_\mathrm{sys}} \int_\mathrm{ray} \nabla_\alpha \,\rho\mathopen{}\left[\mathbf{r}\mathopen{}\left(s\right)\right] \mathrm{d}s.
    \label{equ:deflection:continuous}
\end{equation}
In this expression, $n_0$ is the ambient refractive index, $d$ is the distance from the center of the probe volume to the background, $\psi$ is the pixel pitch, and $C_\mathrm{sys}$ is an overall system constant. The indicator function, $\mathbf{r} \ \colon \mathbb{R}^1 \to \mathbb{R}^3$, maps the distance along a ray, $s$, to the corresponding 3D location, as illustrated above in Fig.~\ref{fig:BOS setup}. Equation~\eqref{equ:deflection:continuous} presumes that deflections are small and the path length through the domain is short compared to $d$. Equivalent expressions are derived in \cite{Atcheson2008, Nicolas2016, Grauer2020}.\par

\subsection{Discrete deflection model}
\label{sec:BOS:discrete}
Algebraic tomography requires a discrete approximation to the forward measurement model, which is inverted by a reconstruction algorithm. To start, the field of interest is discretized using the basis $\Phi = \{\varphi_j\}_{j=1}^N$, in which $\varphi_j$ is the $j$th basis function out of $N$ such functions. Taking this approach, the density field is approximated as follows:
\begin{equation}
    \rho\mathopen{}\left(\mathbf{x}\right) \approx \sum_{j}^N \rho_j \,\varphi_j\mathopen{}\left(\mathbf{x}\right),
    \label{equ:discrete field}
\end{equation}
where $\rho_j$ is a coefficient that scales $\varphi_j$ and the field is represented by the $N\times1$ vector $\boldsymbol\uprho = \{\rho_j\}_{j=1}^N$. Next, the discrete density field is substituted into Eq.~\eqref{equ:deflection:continuous}. Consider the $i$th ray,
\begin{subequations}
    \begin{align}
        \delta_{\alpha,i} &\approx C_\mathrm{sys} \int_\text{ray} \nabla_\alpha \left\{ \sum_{j=1}^N \rho_j \,\varphi_j\mathopen{} \left[\mathbf{r}_i\mathopen{}\left(s\right)\right] \right\} \mathrm{d}s \label{equ:discrete approximation:sequential} \\
        &= C_\mathrm{sys} \sum_{j=1}^N \rho_j
        \underbrace{\int_\text{ray} \nabla_\alpha \,\varphi_j\mathopen{} \left[\mathbf{r}_i\mathopen{}\left(s\right)\right] \mathrm{d}s}_{D_{\alpha,i,j}}, \label{equ:discrete approximation:reversed}
    \end{align}
    \label{equ:discrete approximation}%
\end{subequations}
where $\mathbf{r}_i$ indicates a path along the $i$th ray. Since the integrals over $\varphi_j$ do not depend on $\rho_j$, they can be precomputed to form a deflectometry matrix, $\mathbf{D}_\alpha$, with elements, $D_{\alpha,i,j}$, given by the integration in Eq.~\eqref{equ:discrete approximation:reversed}. This matrix relates the discrete density field, $\boldsymbol\uprho$, to an $M\times1$ vector of $\alpha$-direction deflections, $\boldsymbol\updelta_\alpha = \{\delta_{\alpha,i}\}_{i=1}^M$, such that $\mathbf{D}_\alpha \in \mathbb{R}^{M\times N}$ and
\begin{equation}
    \boldsymbol\updelta_\alpha = C_\mathrm{sys} \,\mathbf{D}_\alpha \boldsymbol\uprho.
    \label{equ:deflection:discrete}
\end{equation}
For axisymmetric flows, we employ a Sipkens deflectometry operator \cite{Sipkens2021}, which is specified in \ref{app:bases and kernels}. For planar fields, we invoke the paraxial assumption and directly approximate Eq.~\eqref{equ:deflection:continuous} by linear ray tracing.\par

\subsection{OF}
\label{sec:BOS:OF}
\begin{figure}[ht]
    \centering
    \includegraphics[width=5.5in]{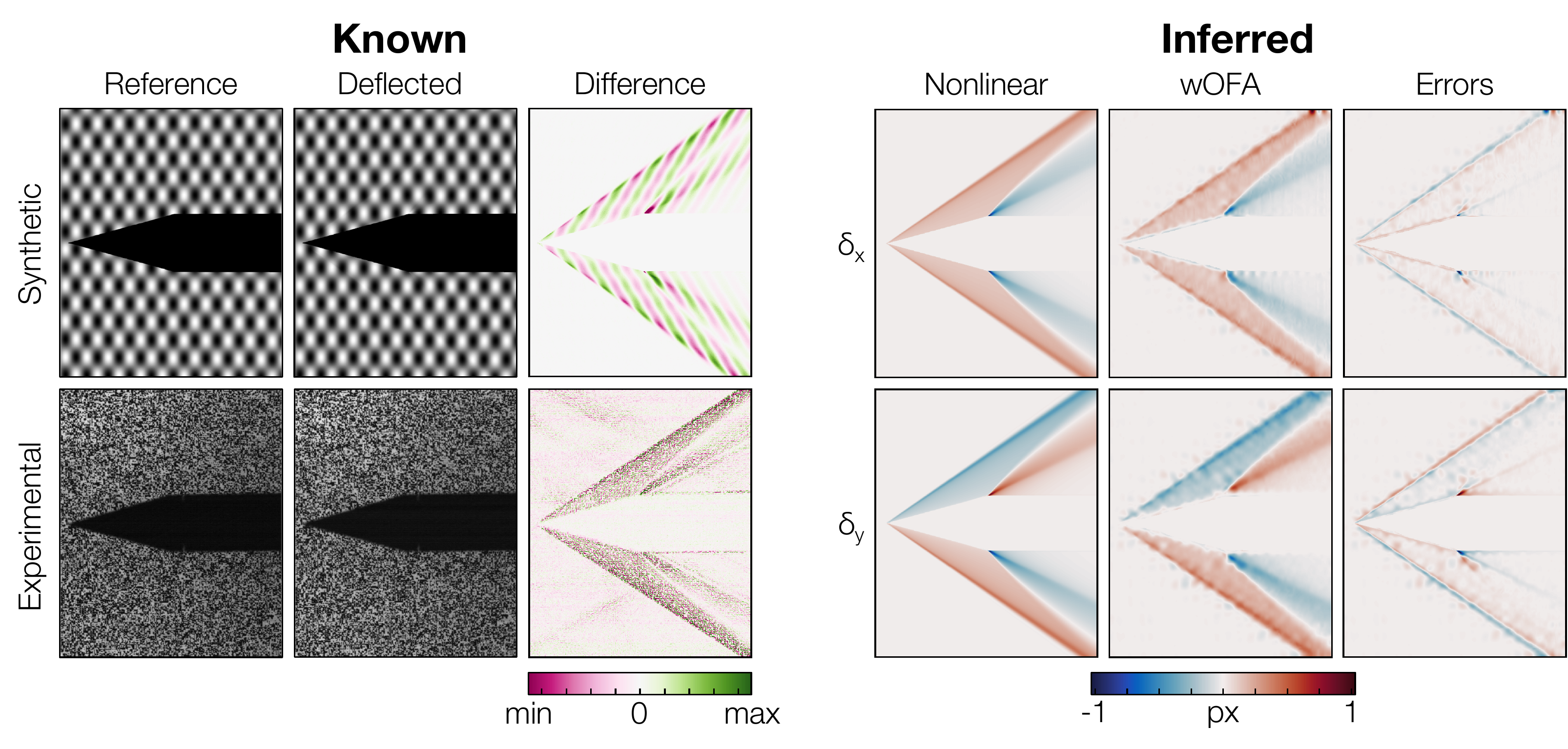}
    \caption{Synthetic and experimental images of a Mach 2 cone cylinder scenario; image differences are processed with an OF algorithm to estimate the deflection field (left). Exact deflections from the synthetic scenario, determined via nonlinear ray tracing, are compared to estimates from an OF algorithm (right).}
    \label{fig:image differences}
\end{figure}

In BOS, refraction results in the \textit{apparent} 2D motion of a background pattern between a reference image, $I_\mathrm{ref}$, and a distorted (or ``deflected'') image, $I_\mathrm{def}$. This corresponds to a classic problem in computer vision known as optical flow, in which the image pair is used to infer a set of deflection vectors \cite{Szeliski2010}. Figure~\ref{fig:image differences} shows synthetic and experimental examples of $I_\mathrm{ref}$ and $I_\mathrm{def}$ for a cone cylinder shock scenario as well as the image differences, $\Delta I = I_\mathrm{def} - I_\mathrm{ref}$. Note that these differences are nearly imperceptible when the images are placed side by side. Nevertheless, the cone shock structure is clearly visible in the plot of $\Delta I$. The image pair is used to infer $\delta_\mathrm{x}$ and $\delta_\mathrm{y}$ at each pixel, which collectively form the measurement vectors for Eq.~\eqref{equ:deflection:discrete}, i.e., $\boldsymbol\updelta_\mathrm{x}$ and $\boldsymbol\updelta_\mathrm{y}$.\par

The transformation of $I_\mathrm{ref}$ into $I_\mathrm{def}$ is modeled in terms of a 2D field of deflection vectors, which warp the original intensity distribution \cite{Davies2004}. There are three key assumptions that underlie OF for BOS \cite{Atcheson2009}:
\begin{enumerate}
    \item Changes in the scene are strictly due to refraction, as opposed to motion or dynamic lighting (shadows, reflections, etc.).
    
    \item The intensity of features is conserved, meaning there is neither emission from the flow nor extinction of light from the background. Stated mathematically,
    \begin{equation}
        I\mathopen{}\left(x, y, t\right) = I\mathopen{}\left(x + \delta_\mathrm{x}, y + \delta_\mathrm{y}, t + \delta_\mathrm{t}\right),
        \label{equ:intensity conservation}
    \end{equation}
    where $\delta_\mathrm{t}$ is the time interval between frames.
    
    \item The magnitude of deflections is small.
\end{enumerate}
Given these assumptions, the OF equation can be approximated by a first-order Taylor series expansion of Eq.~\eqref{equ:intensity conservation}, 
\begin{equation}
    I\mathopen{}\left(x, y, t\right) \approx I\mathopen{}\left(x, y, t\right)
    + \frac{\partial I}{\partial x}\delta_\mathrm{x}
    + \frac{\partial I}{\partial y}\delta_\mathrm{y}
    + \frac{\partial I}{\partial t}\delta_\mathrm{t} + \dots
    \label{equ:Taylor series}
\end{equation}
Methods based on this expression, called gradient-based OF, are invalid for large deflections, e.g., in particle-based OF velocimetry, in which case Eq.~\eqref{equ:intensity conservation} must be solved with a variation algorithm \cite{Schmidt2019, Schmidt2020b}. However, BOS deflections are generally small enough to satisfy the Taylor series expansion \cite{Goldhahn2007}. Time plays an arbitrary role in the OF equations in BOS. Therefore, $\delta_\mathrm{t}$ may be set to unity such that Eq.~\eqref{equ:Taylor series} reduces to
\begin{equation}
    I_x \,\delta_\mathrm{x} + I_y \,\delta_\mathrm{y} = -I_t,
    \label{equ:optical flow}
\end{equation}
where $I_x$ and $I_y$ are finite difference approximations to the partial derivatives in Eq.~\eqref{equ:Taylor series}, evaluated via the reference image, and $I_t$ is set to $\Delta I$. Equation~\eqref{equ:optical flow} applies to each pixel and contains two unknowns, $\delta_\mathrm{x}$ and $\delta_\mathrm{y}$, for each equation (with one equation per pixel).\par

The Horn--Schunck OF closure is ubiquitous since it is both easy to implement and produces acceptable solutions for a wide array of applications \cite{Horn1981}. However, the method amounts to \textit{first-order} Tikhonov regularization, which is known to produce overly-smooth results in the context of flow field measurement \cite{Corpetti2006, Yuan2005, Kadri2013}. Recently, a variational method called wavelet-based OF analysis (wOFA) was shown to yield more accurate deflection fields for BOS than Horn--Schunck OF \cite{Schmidt2021}. In wOFA, the deflection field is represented using a wavelet basis and the (hypothetical) distorted image, $I(x+\delta_\mathrm{x}, y+\delta_\mathrm{y}, t+\delta_\mathrm{t})$, is directly evaluated via bi-cubic spline interpolation; see \cite{Schmidt2020b} for a complete description of the method. This approach allows for the evaluation of residuals from Eq.~\eqref{equ:intensity conservation} without computing the finite differences of $I$, $I_x$ and $I_y$, as must be done in gradient-based OF. Further, the wavelet transform exploits regularities in the deflection field to facilitate a compressed representation of $\boldsymbol\updelta_x$ and $\boldsymbol\updelta_y$, reducing the number of unknowns.\par

We use the wOFA procedure described in \cite{Schmidt2021} for deflection sensing in our conventional BOS workflow. Sample deflections from wOFA applied to the noise-free synthetic cone shock image pair are shown in Fig.~\ref{fig:image differences}. This figure also shows the \textit{exact} deflections that were determined by nonlinear ray tracing. Despite the high-accuracy of the wOFA approach, ``wavelet fingerprints'' can be seen throughout the shocked region. These high-frequency discrepancies between the ideal (true) deflections and wOFA estimates act as measurement errors in the reconstruction algorithm.\par

\subsection{Unified BOS}
\label{sec:BOS:unified}
Until recently, deflection sensing and reconstruction were performed sequentially in BOS. As a result, non-physical regularization was needed in the deflection sensing step because $\boldsymbol\updelta_\mathrm{x}$ and $\boldsymbol\updelta_\mathrm{y}$ are not directly related to the density field (rather, they are LoS-integrated quantities). However, Grauer and Steinberg \cite{Grauer2020} observed that coupled gradient OF equations, i.e., one instance of Eq.~\eqref{equ:optical flow} for each pixel, can formulated as a matrix system. The discrete BOS model from Sect.~\ref{sec:BOS:discrete} can then be substituted into the matrix OF equation to relate the density field to raw image difference measurements. In other words, the density field can be reconstructed from a vector of image difference data in a single step. All regularization in this ``unified'' procedure is applied to the density field per se, and the errors associated with deflection sensing are avoided.\par

\begin{figure}[ht]
    \centering
    \includegraphics[width=5.25in]{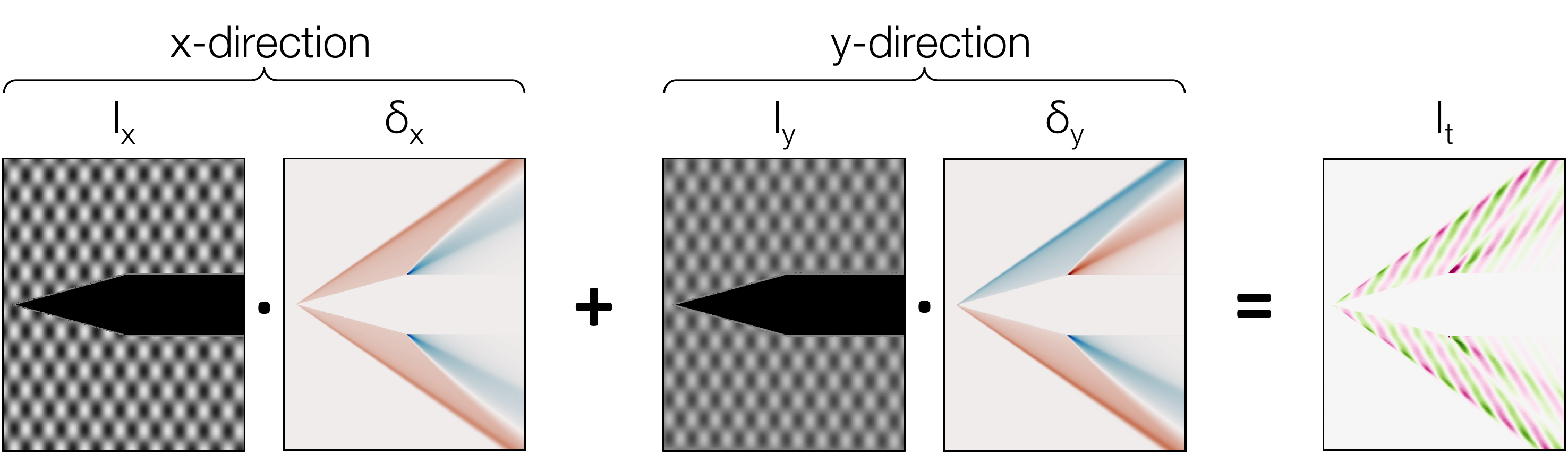}
    \caption{Visualization of unified BOS: $x$- and $y$-direction image gradients are multiplied by deflections to estimate the image difference, i.e., Eq.~\eqref{equ:optical flow} is calculated at each pixel. Color scales are the same as in Fig.~\ref{fig:image differences}.}
    \label{fig:unified BOS}
\end{figure}

Figure~\ref{fig:unified BOS} is a visualization of unified BOS in matrix form. Equation~\eqref{equ:optical flow} is applied to each pixel, independently. This amounts to a diagonal matrix operation applied to $\boldsymbol\updelta_\mathrm{x}$ and $\boldsymbol\updelta_\mathrm{y}$, which may be expressed as a function of $\boldsymbol\uprho$. To start, diagonal matrices are formed from the horizontal and vertical intensity gradients, $\mathbf{X}_{i,i} = I_{x,i}$ and $\mathbf{Y}_{i,i} = I_{y,i}$ for $i = 1, 2, \dots, M$ for a system of $M$ pixels. In this work, we employ second-order central differences to compute $I_x$ and $I_y$. Next, we construct a data vector from the image pair, $\mathbf{b} = \{-I_{t,i}\}_{i=1}^M$. Lastly, the $x$- and $y$-direction instances of Eq.~\eqref{equ:deflection:discrete} are substituted in for $\boldsymbol\updelta_\mathrm{x}$ and $\boldsymbol\updelta_\mathrm{y}$, respectively, to relate $\mathbf{b}$ to $\boldsymbol\uprho$,
\begin{equation}
    \mathbf{X}\boldsymbol\updelta_\mathrm{x} + \mathbf{Y}\boldsymbol\updelta_\mathrm{y} =  \underbrace{C_\mathrm{sys} \left(\mathbf{X}\mathbf{D}_\mathrm{x} + \mathbf{Y}\mathbf{D}_\mathrm{y} \right)}_\mathbf{A}\boldsymbol\uprho = \mathbf{b}.
    \label{equ:unified BOS}
\end{equation}
Here, $\mathbf{A}$ is the $M\times N$ unified BOS operator. Equation~\eqref{equ:unified BOS} can be solved with the same reconstruction techniques developed to solve Eq.~\eqref{equ:deflection:discrete}.\par

\subsection{Axisymmetric reconstruction}
\label{sec:BOS:reconstruction}
There are many methods for reconstructing a 2D or 3D flow field from path-integrated measurements, as recently reviewed in \cite{Grauer2023}. This paper considers the application of BOS to axisymmetric and planar flows. The geometric simplicity of these configurations facilitates an efficient representation of the fields and a simplified measurement model. A brief overview of Abel inversion for axisymmetric BOS and a regularization technique for unified BOS are presented below.\par

\subsubsection{Abel inversion for BOS}
\label{sec:BOS:reconstruction:Abel}
Path integrals through a radially symmetric object are described by the Abel transform, which has an explicit, analytical inverse. The forward transform \textit{loosely} corresponds to the measurement model for many tomography modalities, such as absorption or emission tomography. Consequently, the inverse Abel transform can be adapted to reconstruct absorbance or emission data when the target object is axisymmetric. However, the BOS measurement model features a path integral over \textit{gradients} of the density field, as opposed to integrals over $\rho$ per se. This complication necessitates a tailored reconstruction strategy.\par

Recall the forward model for radial ($y$-axis) deflections in Eq.~\eqref{equ:deflection:continuous},
\begin{equation}
    \delta_\mathrm{y} = C_\mathrm{sys} \int_\mathrm{ray} \nabla_y \,\rho\mathopen{}\left(s\right) \mathrm{d}s
    = C_\mathrm{sys} \nabla_y \underbrace{\int_\mathrm{ray} \rho\mathopen{}\left(s\right) \mathrm{d}s}_{\bar\rho}
    = C_\mathrm{sys} \nabla_y \,\bar\rho,
    \label{equ:projected density}
\end{equation}
where $\bar\rho$ is the so-called projected density field, which corresponds to the forward Abel transform of $\rho$ when the flow is axisymmetric \cite{Raffel2015}. The gradient and integral operations in Eq.~\eqref{equ:projected density} are reversible,\footnote{This result is a consequence of the Leibniz integral rule, assuming constant bounds of integration.} leading to two distinct methods for reconstruction, which are outlined in Fig.~\ref{fig:BOS workflow}.
\begin{enumerate}
    \item Indirect reconstruction: Eq.~\eqref{equ:projected density} is used to construct a Poisson equation that may be solved for $\bar\rho$, which is recovered via the standard inverse Abel transform.\footnote{These steps are commonly switched in 3D BOS tomography \cite{Atcheson2008}; this is also possible in axisymmetric BOS (see the work of Ota et al. \cite{Ota2015, Hirose2019}, for instance), but doing so is rare.}
    
    \item Direct reconstruction: the Abel transform is modified to incorporate $\nabla_y$ and $\rho$ is directly determined from $\delta_y$. This resultant transform is often referred to as the ``deflectometry'' version \cite{Kolhe2009}.
\end{enumerate}
The Poisson equation in indirect Abel inversion contains gradients of $\delta_\mathrm{x}$ and $\delta_\mathrm{y}$. Differentiating these variables amplifies errors from the deflection sensing procedure. Direct methods are comparatively stable, so we adopt the latter approach in this work.\par

The deflectometry Abel transform and its inverse are
\begin{subequations}
    \begin{align}
        \delta_\mathrm{y}\mathopen{}\left(y\right) &= 2 \,C_\mathrm{sys} \int_{y}^R \nabla_y \,\rho\mathopen{}\left(r\right) \frac{y}{\sqrt{r^2 - y^2}} \,\mathrm{d}r \quad\text{and}
        \label{equ:deflectometry transform:forward} \\
        \rho\mathopen{}\left(r\right) &= -\frac{1}{\pi \,C_\mathrm{sys}} \int_r^R \delta_\mathrm{y}\mathopen{}\left(y\right) \frac{1}{\sqrt{y^2 - r^2}} \,\mathrm{d}y,
        \label{equ:deflectometry transform:inverse}
    \end{align}
    \label{equ:deflectometry transform}%
\end{subequations}
where the radial direction is aligned with the $y$-axis and $R$ is the outer radius of the flow, beyond which $\delta_\mathrm{y}$ is zero \cite{Kolhe2009}. Notice that Eq.~\eqref{equ:deflectometry transform} requires continuous deflection data, which is not available in practice, so the expression must be discretized. The resultant direct deflectometry Abel inversion for a discrete field is
\begin{equation}
    \boldsymbol\uprho = C_\mathrm{sys}^{-1} \,\mathbf{K} \boldsymbol\updelta_\mathrm{y},
    \label{equ:discrete Abel transform}
\end{equation}
where $\mathbf{K}$ is an inverse Abel operator that is obtained by discretizing Eq.~\eqref{equ:deflectometry transform:inverse}. Since $\delta_\mathrm{y}$ is resolved at discrete intervals (viz., elements of $\boldsymbol\updelta_\mathrm{y}$), $\mathbf{K}$ effectively interpolates this data. The two most common techniques for building $\mathbf{K}$ follow Simpson's 1/3 rule and the two-point method. Expressions for the elements of $\mathbf{K}$ derived using Simpson's 1/3 rule and two-point interpolation are provided in \ref{app:bases and kernels}.\par

Unfortunately, the simplifications required to obtain $\mathbf{K}$, such as the assumption of parallel rays, are often violated by practical imaging systems. Further, Eq.~\eqref{equ:discrete Abel transform} is a 1D transform that must be independently applied to each axial segment of a 2D flow. Neither axial gradients in the flow nor axial deflections are considered in the inversion. Therefore, inverting a high-fidelity 2D forward model like the unified BOS matrix is more robust \cite{Daun2006, Sipkens2021}.\par

\subsubsection{Tikhonov regularization}
\label{sec:reconstruction:Tikhonov}
The unified BOS model relates the (as yet unknown) density field to a measured image difference vector, $\mathbf{b}$. Therefore, this model must be inverted to estimate $\boldsymbol\uprho$ from $\mathbf{b}$. This is a discrete ill-posed inverse problem and additional information is required for stability \cite{Daun2016}. We utilize a \textit{second-order} Tikhonov penalty to promote smooth solutions:
\begin{equation}
    \boldsymbol\uprho_\omega = \mathrm{arg}\,\underset{\boldsymbol\uprho}{\mathrm{min}} \left( \left\lVert\mathbf{A}\boldsymbol\uprho - \mathbf{b}\right\rVert_2^2 + \omega_\mathrm{Tik}^2 \left\lVert\mathbf{L}\boldsymbol\uprho \right\rVert_2^2 \right).
    \label{equ:Tikhonov solution}
\end{equation}
In this expression, $\mathbf{L}$ is a discrete Laplacian operator (or ``Tikhonov matrix''), defined in \ref{app:bases and kernels}, $\omega_\mathrm{Tik}$ is a regularization parameter that controls the influence of the penalty, and solutions may be obtained with a linear least squares algorithm. As $\omega_\mathrm{Tik}$ goes to zero, $\boldsymbol\uprho_\omega$ approaches the least squares solution, which is highly sensitive to noise in $\mathbf{b}$. Conversely, for very large values of $\omega_\mathrm{Tik}$, Eq.~\eqref{equ:Tikhonov solution} is minimized by the uniform vector that minimizes the measurement residuals. At moderate values of $\omega_\mathrm{Tik}$, however, $\boldsymbol\uprho_\omega$ corresponds to a smooth field that approximately satisfies the measurement equations. Therefore, we optimize the regularization parameter through a phantom study, wherein $\boldsymbol\uprho_\omega$ is computed from a synthetic image pair and compared to the exact density distribution. This process is repeated for a large range of regularization parameters, and we set $\omega_\mathrm{Tik}$ to the value which maximizes the accuracy of $\boldsymbol\uprho_\omega$.\par

\section{Physics-informed BOS}
\label{sec:physics-informed BOS}
Current BOS algorithms can produce a quantitative estimate of the density field in a target flow, but regularization is required at each stage of the workflow. Although most regularization schemes for BOS are \textit{inspired} by physics, existing methods are ultimately incompatible with the underlying flow fields and thus give rise to errors or ``reconstruction artifacts.''\par

Data assimilation (DA) is a promising alternative to the standard suite of BOS techniques. DA algorithms seek to solve the equations governing fluid motion subject to data-based constraints \cite{Hayase2015}. This approach avoids the pitfalls of the BOS methods described above and also provides access to the latent velocity, pressure, and energy fields. There are numerous methods that can be used to solve (or approximately solve) the governing equations while conforming to experimental measurements. Kalman filter \cite{Cornick2009, Ali2022}, state observer \cite{Saredi2021}, adjoint--variational \cite{Mons2021, Wang2022b}, and hybrid simulation \cite{Vinnichenko2022} algorithms have all been used to reconstruct flow fields with input from an experiment. For instance, local ensemble Kalman filter DA was employed to forecast temperature and velocity fields in a Rayleigh--B\'enard convection cell from a set of experimental shadowgraphs \cite{Cornick2009}. A similar framework was developed at ONERA by Ali et al. \cite{Ali2022}, who repeatedly solved the RANS equations and sequentially updated turbulence model parameters with a Kalman filter to match synthetic BOS measurements. State observer methods incorporate proportional or proportional--integral feedback, based on measurements of one or more fields, into the governing equations \cite{Saredi2021}. Variational techniques optimize a control vector, such as the initial flow state, to minimize an arbitrary data loss \cite{Mons2021, Wang2022b}, and hybrid CFD simulations are conducted with one or more fields or parameters that are fixed by data. As an example of the latter technique, Vinnichenko et al. \cite{Vinnichenko2022} conducted a hybrid simulation of natural convection using a BOS-based estimate of the temperature field to determine the buoyancy term.\par

Unfortunately, these DA methods come at a high computational cost. Furthermore, all examples of BOS DA reported to-date have employed either (a) idealized synthetic BOS data \cite{Ali2022} or (b) temperature field estimates from a conventional BOS algorithm \cite{Cai2021, Vinnichenko2022}. When the data loss, constraint, or forcing term in a DA algorithm includes \textit{reconstructions}, as opposed to the measurement model and raw signal, the resultant fields are adversely affected by non-physical reconstruction artifacts introduced by the BOS algorithm. These artifacts can bias the DA algorithm or even prevent convergence, so it is essential to accurately mimic the image formation process in the data loss term \cite{Molnar2022}.\par

We conduct physics-informed BOS using a low-cost, flexible, easy-to-implement DA scheme in which the flow is represented with a PINN. The goal is to optimize an aggregate loss, consisting of data and physics residuals, using mature deep learning tools. The result is accurate, spatially-resolved estimates of all the flow fields.\par

\subsection{PINN framework}
\label{sec:physics-informed BOS:framework}
\begin{figure}[ht]
    \centering
    \includegraphics[width=5.75in]{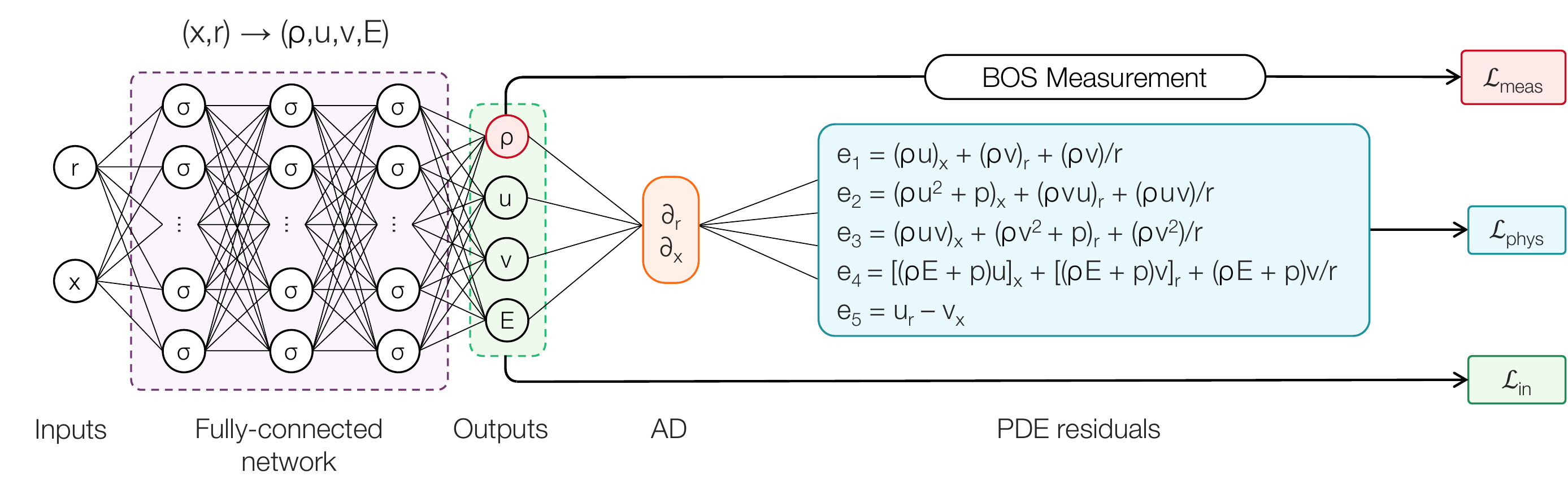}
    \caption{Our PINNs map axial and radial (or vertical) coordinates to the fields of interest. The unified BOS measurement model is embedded in the data loss term; the Euler and irrotationality equations are employed for physics; experimental wind tunnel conditions are enforced at the inlet.}
    \label{fig:PINN architecture}
\end{figure}

Physics-informed neural networks utilize a deep, feedforward network to map spatio-temporal inputs to flow fields \cite{Raissi2019}. Figure~\ref{fig:PINN architecture} shows the architecture of a PINN set up for physics-informed BOS. We utilize $(x,r)$ and $(x,y)$ as inputs for axisymmetric and planar flows, respectively. Density ($\rho$), velocity ($u$ and $v$), and total energy ($E$), are the outputs. Automatic differentiation (AD) is employed to calculate exact partial derivatives of the network, and these partials are used to evaluate the governing equations throughout the measurement domain. The PINN does not generally satisfy these equations and the residuals are added up in a physics loss. Separately, a data (or measurement) loss is obtained by evaluating the unified BOS measurement model and comparing synthetic image differences, computed using the density field outputted by the PINN, to experimental image differences. The total objective loss (data + physics) is minimized via backpropagation to estimate the flow field in functional form.\par

In this work, we specify a physics loss using the compressible, steady, 2D Euler equations. This is an appropriate choice for the present demonstration, which features invisid flow, but viscous effects, transience, and 3D fields can be included as necessary. For now, each of the non-dimensional Euler equations is re-arranged to isolate a residual, $\varepsilon$:
\begin{subequations}
    \begin{align}
        \varepsilon_1 &= \left(\rho \,u\right)_x + \left(\rho \,v\right)_r + \beta\left(\frac{1}{r}\rho \,v\right) \label{equ:Euler residuals:mass}\\
        \varepsilon_2 &= \left(\rho \,u^2 + p\right)_x + \left(\rho \,v \,u\right)_r + \beta\left(\frac{1}{r}\rho \,u \,v\right) \label{equ:Euler residuals:z momentum}\\
        \varepsilon_3 &= \left(\rho \,u \,v\right)_x + \left(\rho \,v^2 + \,p\right)_r + \beta\left(\frac{1}{r}\rho \,v^2\right) \label{equ:Euler residuals:r momentum}\\
        \varepsilon_4 &= \left[\left(\rho \,E + p\right) u\right]_x +  \left[\left(\rho \,E + p\right) \,v\right]_r + \beta\left[\frac{1}{r}\left(\rho \,E + p\right) v\right]. \label{equ:Euler residuals:energy}
    \end{align}
    \label{equ:Euler residuals}%
\end{subequations}
Here, $(\cdot)_x$ and $(\cdot)_r$ denote partial derivatives with respect to $x$ and $r$ (naturally, $r$ is replaced with $y$ for planar cases) and $\beta$ is the radial source coefficient which equals 1 for axisymmetric flow and 0 for planar flow. In order to compute residuals along the axis of symmetry, we multiply the right side of Eq.~\eqref{equ:Euler residuals} by $r$. The field variables in Eq.~\eqref{equ:Euler residuals} must be scaled by an appropriate reference to obtain dimensional values. We adopt the inflow conditions as our reference set, with a single reference for both components of velocity. A polytropic equation of state is employed to close the Euler equations,
\begin{equation}
        p = \left(\gamma - 1\right) \rho \left[E - \frac{1}{2} \left(u^2 + v^2\right)\right],
        \label{equ:state equation}
\end{equation}
where $\gamma$ is the ratio of specific heats. Lastly, since this paper is concerned with approximately inviscid flow and approximately uniform inlet conditions, we consider an additional equation that is satisfied by irrotational velocity fields,
\begin{equation}
    \varepsilon_5 = u_r - v_x.
    \label{equ:irrotational flow}
\end{equation}
Residuals from Eqs.~\eqref{equ:Euler residuals} and \eqref{equ:irrotational flow} are integrated over the domain, culminating in an overall physics loss,
\begin{equation}
    \mathcal{L}_\mathrm{phys} = \frac{1}{\pi R^2 L} \int_0^L \int_0^R \left\lVert \left[\varepsilon_1,  \dots, \varepsilon_5\right] \right\rVert_2^2 2\pi r \,\mathrm{d}r \,\mathrm{d}x,
    \label{equ:physics residual}
\end{equation}
where $L$ and $R$ are the length and radius of the measurement domain. (Once again, the appropriate modifications are made for planar cases.) This expression is approximated by Monte Carlo sampling, and residuals from the irrotational equation are omitted in certain cases, as explicitly noted in the results sections.\par

There is an infinite set of flow fields that minimizes Eq.~\eqref{equ:physics residual}. Therefore, we connect the PINN to our experimental target with a data loss. This loss features the unified BOS model, such that the density field from the PINN is employed to compute image differences, which may then be compared to the unprocessed experimental images. In other words, our data loss is
\begin{equation}
    \mathcal{L}_{\mathrm{meas}} = \left\lVert \mathbf{A} \boldsymbol\uprho - \mathbf{b}  \right\rVert_2^2, 
    \label{equ:data loss}
\end{equation}
where $\mathbf{b}$ is contains experimental image differences, $\mathbf{A}$ is the unified BOS operator, and $\boldsymbol\uprho$ is a vector of density data that is obtained from the PINN. We query the PINN at the support nodes of our basis, $\Phi$, to construct $\boldsymbol\uprho$. It should also be emphasized that the PINN outputs normalized values; hence, the outputted density data must be multiplied by a reference density (the inlet density in our case) to populate $\boldsymbol\uprho$.\par

The final loss term is an inlet boundary condition, which is generally known to first-order for experiments such as the wind tunnel tests described in Sect.~\ref{sec:scenarios}. Our inlet loss is
\begin{equation}
    \mathcal{L}_\mathrm{in} = \frac{1}{\pi R^2}  \int_0^R \left\lVert \left[\rho-\rho_\mathrm{in}, u - u_\mathrm{in}, v-v_\mathrm{in}, E-E_\mathrm{in}\right] \right\rVert_2^2 2 \pi r \,\mathrm{d}r,
    \label{equ:inlet residual}
\end{equation}
where $\rho$, $u$, $v$, and $E$ are evaluated at $x = 0$ and $(\cdot)_\mathrm{in}$ denotes an inlet value and the radial integral is replaced with a linear one for planar flow. The objective loss to be minimized is
\begin{equation}
    \mathcal{L}_\mathrm{total} = \omega_\mathrm{meas}\,\mathcal{L}_\mathrm{meas} + \omega_\mathrm{phys}\,\mathcal{L}_\mathrm{phys} + \omega_\mathrm{in}\,\mathcal{L}_\mathrm{in}.
    \label{equ:objective loss}
\end{equation}
This equation is uniquely minimized by the true flow fields so long as the PINN is big enough to express the flow, but the relative weight of the loss terms can help or hamper training. We optimize the weighting parameters, $\omega_\mathrm{phys}$, $\omega_\mathrm{meas}$, and $\omega_\mathrm{in}$, by conducting a simple parameter sweep with a synthetic case; optimal weights from this test, $\omega_\mathrm{meas}/\omega_\mathrm{phys} = 10$ and $\omega_\mathrm{in}/\omega_\mathrm{phys} = 100$, are used throughout the paper. Several groups have developed heuristics to programmatically assign weights to individual components of a PINN's objective loss, e.g., \cite{Wang2022a, Wang2021, Jin2021}. However, the benefits of these adaptive techniques are generally marginal in the presence of realistic measurement noise, as discussed in \cite{Molnar2022}. Another promising approach is to use a traditional algorithm for constrained optimization, like the \textit{alternating direction method of multipliers}, which was applied to PINNs by Basir et al. \cite{Basir2022}. We plan to assess this scheme in future research.\par

\subsection{Network architecture and training}
\label{sec:physics-informed BOS:architecture}
The PINNs used for this work are implemented in TensorFlow \cite{Abadi2016}. Networks that represent planar flows comprise five hidden layers, with 50 neurons per output variable, while the networks representing axisymmetric flows have ten hidden layers to ensure adequate expressivity. We employ swish activation functions unless otherwise noted. Weights are randomly initialized with a standard normal distribution and biases are initially set to zero.\par

Training is performed by minimizing Eq.~\eqref{equ:objective loss} with the Adam optimizer \cite{Kingma2014} at a learning rate of $10^{-3}$ for the first three passes through the full image dataset and $10^{-4}$ thereafter. The PINNs are trained until the total loss plateaus; we define a plateau as a 5000-iteration stretch over which the 500-iteration running average of $\mathcal{L}_\mathrm{total}$ varies by less than 0.5\%. Here, an iteration is defined as a single update of the parameters by the Adam optimizer. Reconstructions are computed on an NVIDIA Tesla P100 graphics processing unit. Planar reconstructions initially took five hours, on average, and axisymmetric reconstructions took around nine hours; optimization of the code in TensorFlow~2.9.2 has reduced these times to approximately two and three hours, respectively.\par

\subsection{PINNs for high-speed flow}
\label{sec:physics-informed BOS:hyperbolic}
Supersonic flow is compressible and subject to shock formations in most practical scenarios. The Euler equations are hyperbolic and shocks manifest as discontinuities in the corresponding solution. It is challenging to account for these effects with a numerical solver, which has led to a rich literature on bespoke CFD methods for supersonic conditions. Flow with shocks is similarly problematic for PINNs and several groups are actively adapting traditional CFD techniques to improve the ability of PINNs to represent high-speed flow. We discuss techniques for applying PINNs to hyperbolic equations in \ref{app:physics-informed BOS hyperbolic}.\par 

The use of PINNs to learn real compressible flow fields from experimental data is of particular interest, and several researchers have attempted to simulate this process with a \textit{pseudo-schlieren} scenario. These tests use a loss consisting of point-wise $\boldsymbol\nabla \rho$ ``measurements'' as opposed to LoS-integrated signals. Mao et al. \cite{Mao2020} pioneered the use of a pseudo-schlieren loss to estimate 1D shock-laden airflow (Mao also developed a forward solver for a 2D oblique shock with no schlieren-type data). Cai et al. \cite{Cai2022} reconstructed a synthetic 2D bow shock in the same way, quickly followed by the paper of Jagtap et al. \cite{Jagtap2022}, who utilized domain decomposition (via an extended PINN) to facilitate the representation of oblique and bow shocks as well as an expansion fan. All three studies utilized a dense array of noise-free, synthetic, multi-modal measurements. For instance, Jagtap and coworkers specified 700 local density gradient pairs and 50 pressure taps in their data loss term. By way of context, well-instrumented cones support up to 16 taps \cite{Casper2016}, and a typical field-ready design has no taps at all. Since real schlieren data is LoS-integrated and camera rays may diverge in the measurement volume \cite{Walsh2000, Sipkens2021}, particularly when access windows are required for imaging, we restrict our tests to \textit{plausible} synthetic data and real experimental images.\par

\section{Measurement scenarios}
\label{sec:scenarios}
\begin{figure}[ht]
    \centering
    \includegraphics[width=4.75in]{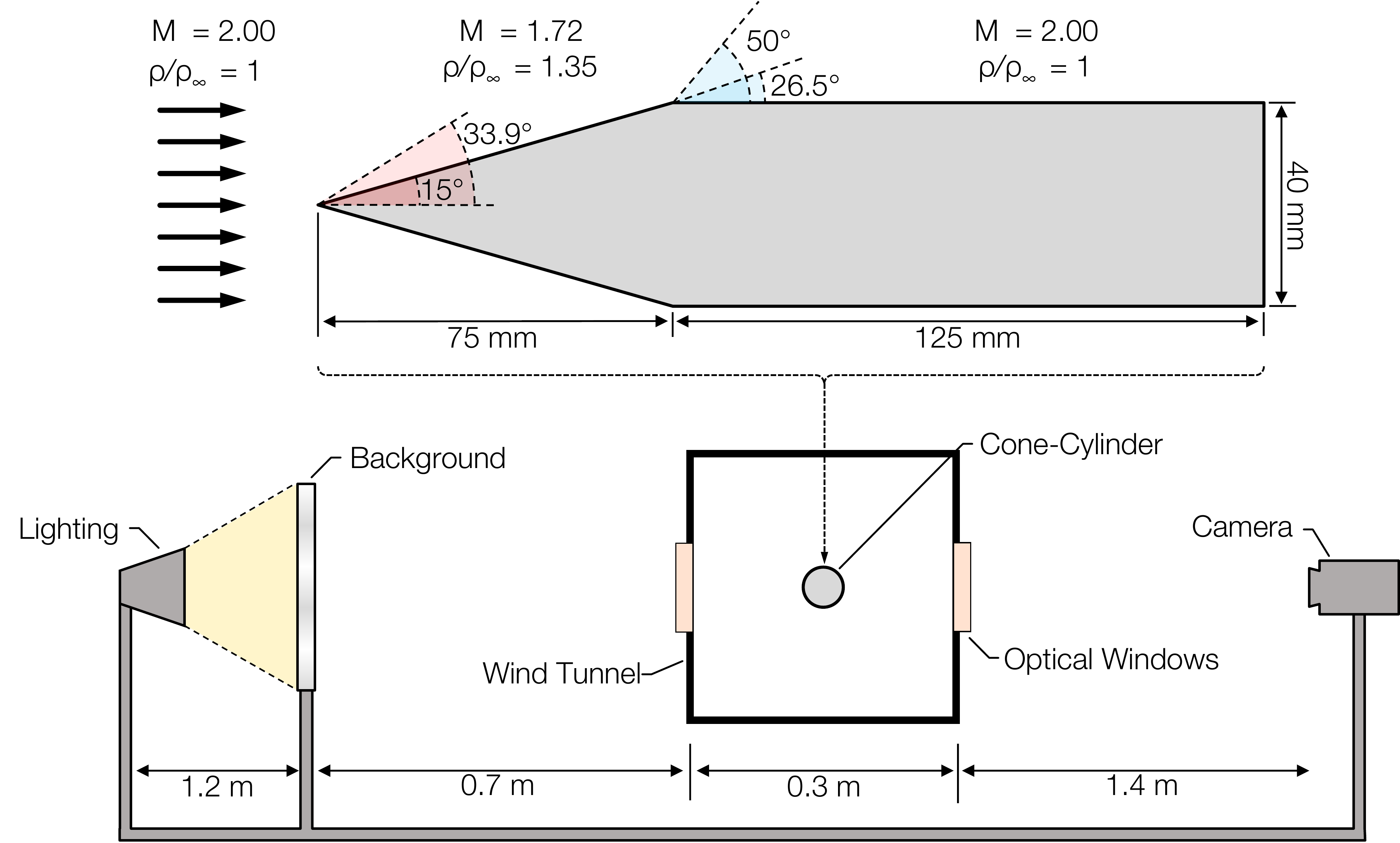}\vspace{0.3em}
    \caption{Schmatic of the cone cylinder experiment at NAL, Bangalore. Mach 2 airflow passes over a 15$^\circ$ cone cylinder. The cylinder is located in the middle of a square $0.3\times0.3$~m$^2$ test section. Photos of a backlit background pattern are recorded through 1''-thick quartz acccess windows using a 1~MP camera.}
    \label{fig:inlet cone schematic}
\end{figure}

We use physics-informed BOS to process experimental and synthetic image data for the axisymmetric cone cylinder scenario depicted in Fig.~\ref{fig:inlet cone schematic}. Additionally, we demonstrate our technique on an analytical (planar) Prandtl--Meyer expansion fan. Both scenarios feature Mach 2 flow, leading to significant shock formations.\par

\subsection{BOS measurements}
\label{sec:scenarios:BOS}

\subsubsection{Experimental setup}
\label{sec:scenarios:BOS:experiment}
The BOS setup featured in this paper is located at the National Aerospace Laboratories (NAL), Bangalore, and is described in detail in \cite{Venkatakrishnan2004}. The facility houses a trisonic wind tunnel and the experiment is conducted in a square test section having a $0.3\times0.3$~m$^2$ profile. Square access windows, approximately 1'' thick, are installed on opposite sides of the tunnel to enable imaging of the background pattern. The model is a 15$^\circ$ half-angle cone cylinder, positioned in the middle of the tunnel with zero inclination. This configuration was chosen because the resultant density field about the cone can be determined from classical cone shock tables \cite{Sims1964}. Testing was conducted for Mach 2 inflow, which we assumed to be uniform throughout the cross section.\par

\begin{figure}[h]
    \centering
    \includegraphics[width=4.75in]{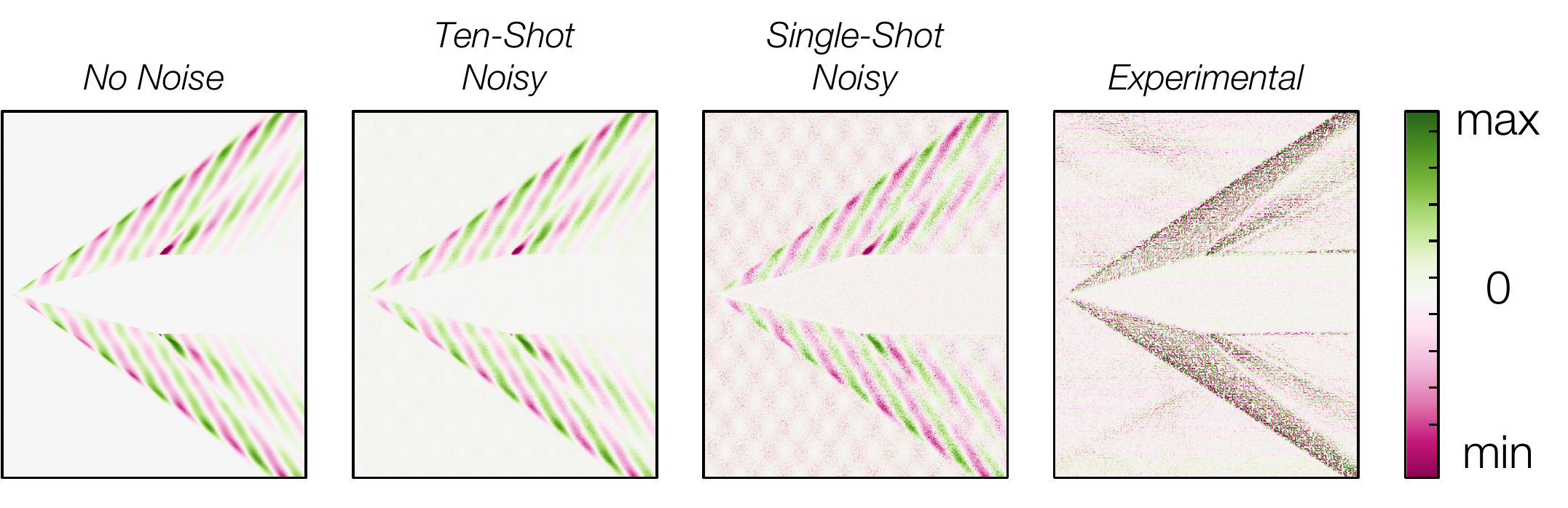}\vspace{0.3em}
    \caption{Image difference data for a Mach 2 cone cylinder experiment. From left to right, the panel depicts clean, ten-shot average, and single-shot synthetic data followed by an experimental acquisition.}
    \label{fig:data panel}
\end{figure}

The background was placed 0.7~m from the tunnel window. Imaging was conducted with a scientific-grade camera (Kodak ES 1.0) having a 1~MP sensor and 9~$\upmu$m pixels. The camera was fitted with a 50~mm lens, stopped down to $f/8$, and placed 1.4~m away from the tunnel, pointing towards the background plate. We consider a $540\times652$~px subsection of the sensor that is centered on the axis of symmetry. Experimental images of the cone are shown in Fig.~\ref{fig:image differences}, and the difference data can be seen on the right side of Fig.~\ref{fig:data panel}. Note that we used a \textit{single pair} of experimental images. Significant artifacts are visible in resulting image differences, which exhibit noticeable striations that are emblematic of fixed pattern noise.\par

\subsubsection{Generating synthetic data}
\label{sec:scenarios:synthetic}
In order to benchmark the accuracy of our reconstruction scheme, we test the method on synthetic images that correspond to ground truth (``phantom'') density fields, introduced in the next section. To start, a pinhole camera model, based on the parameters presented above, is used to determine the principal ray for each point on the sensor; we supersample the sensor at four times the native resolution, i.e., using four rays per pixel. Linear ray tracing is employed for regions of constant density: outside the wind tunnel, inside the tunnel for the reference image, and through the windows (including Snell's Law-type refraction at each interface). Further, we sample the aperture using the method of Cook \cite{Cook1984} to account for the blurring and finite depth-of-field produced by a real aperture. Our measurement operator is thus effectively a ``cone beam'' sensitivity matrix that can mimic BOS measurements recorded with distinct $f$-numbers.\par

While reference images are generated by linear ray tracing throughout the domain, distorted images require nonlinear ray tracing inside the wind tunnel and linear ray tracing outside. Nonlinear ray tracing involves numerically solving the ray equation for a nonuniform refractive index field. We use the fourth-order Runge--Kutta scheme developed by Sharma et al. \cite{Sharma1982} to solve Eq.~\eqref{equ:ray equation}, described in \ref{app:ray tracing}. Lastly, we corrupt some of the synthetic images with noise. Three cases are considered: clean (noise-free) image pairs, single-shot noisy image pairs, and ten-shot average image pairs. Individual noisy images are generated by applying a Gaussian blur filter with a standard deviation of 1.5~px and simulating shot-noise. The latter effect is modeled as a Poisson random variable, assuming a maximum signal of 5000 counts \cite{Foi2008}. Clean, ten-shot noisy, and single-shot noisy difference data can be seen in the first three panels of Fig.~\ref{fig:data panel}. In separate tests, we added 1--10\% centered Gaussian errors to each \textit{noisy} shot to mimic low-to-high levels of thermal noise.\par

Notice that the experimental images in Fig.~\ref{fig:image differences} contain a dot pattern background whereas the synthetic background contains a superposition of axis-aligned sinusoidal waves. Historically dot patterns, inspired by PIV, were used to facilitate cross-correlation deflection sensing. However, since OF takes advantage of intensity gradients in the background, any blank space between the dots is of no use. Therefore, we stick to the sine wave pattern, which has been shown to improve performance \cite{Grauer2020, Schmidt2021}.\par

\subsection{Phantom flow fields}
\label{sec:scenarios:phantoms}

\subsubsection{Axisymmetric cone cylinder shock}
\label{sec:scenarios:phantoms:cone}
We simulate the Mach 2 cone cylinder experiment reported in \cite{Venkatakrishnan2004} to estimate the density, velocity, and energy fields corresponding to our experimental data. Simulations are performed using the compressible Euler solver in SU2~7.3.0 \cite{Economon2016}. The axisymmetric computational domain has a radius of 0.15~m and length of 0.25~m, excluding the cone cylinder geometry. Inlet conditions are specified to match experimental conditions for a settling pressure and temperature of 2.1~kg/cm$^2$ and 300~K, respectively, and $M = 2$, where $M$ is the Mach number. The ratio of specific heats for air is set to the standard value of $1.4$. Discretization errors were confirmed to be minimal through a grid convergence study, which is reported in \ref{app:phantoms}, along with additional details about the simulation.\par

\subsubsection{Planar expansion fan}
\label{sec:scenarios:phantoms:fan}
\begin{figure}[ht]
    \centering
    \includegraphics[width=2.75in]{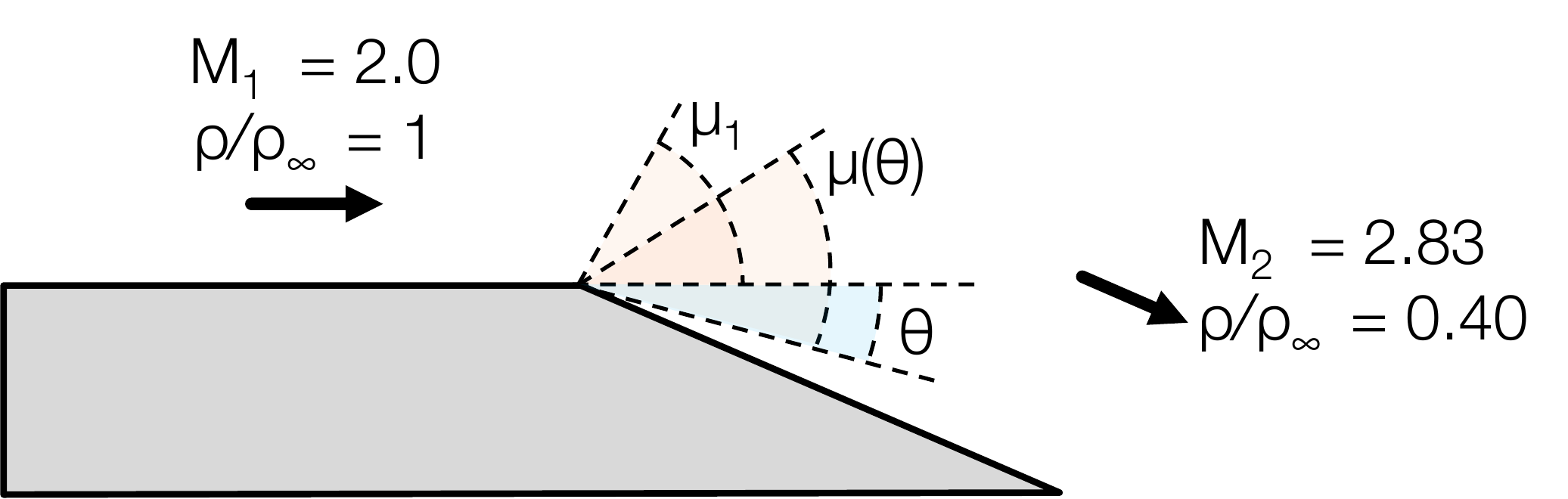}
    \caption{Supersonic flow over a convex ramp leads to an expansion fan. The fan is characterized throughout in terms of a turning angle, $\theta$, which is related to the Mach angle, $\mu$.}
    \label{fig:expansion fan}
\end{figure}

A secondary, analytical phantom is considered, namely: the canonical Prandtl--Meyer expansion fan. This phenomenon occurs when supersonic flow encounters a convex ramp, as illustrated in Fig.~\ref{fig:expansion fan}; the flow accelerates over the ramp and sustains a continuous isentropic expansion across an infinite sequence of ``Mach waves'' \cite{Anderson1990}. We generate an expansion fan for $M_1 = 2$ inflow that encounters a 20$^\circ$ corner. Key parameters of the flow are provided in Table~\ref{tab:fan parameters}, and details about the field calculations are given in \ref{app:phantoms}. When simulating BOS measurements of the fan, we assume a slightly narrower test section of width 3.75~cm to limit the magnitude of deflections, such that the paraxial assumption holds true. A single pair of supersampled images are used in the inversion. Additionally, we consider a few cases with ``pressure tap'' information using the analytical value at the ramp.\par

\begin{table}[ht]
    \renewcommand{\arraystretch}{1.25}
    \caption{Parameters for a Prandtl--Meyer Expansion Fan Phantom}
    \centering
    \begin{tabular}{r c c c c c}
        \hline\hline
        \multicolumn{1}{c}{Region} & \multicolumn{1}{c}{$M$} & \multicolumn{1}{c}{$\rho$, kg/m$^3$} & \multicolumn{1}{c}{$u$, m/s} & \multicolumn{1}{c}{$v$, m/s} & \multicolumn{1}{c}{$p$, kPa}\\
        \hline
        Inflow & $2$ & $0.55$ & $517.6$ & $0.0$ & $26.3$ \\
        Outflow & $2.83$ & $0.22$ & $572.5$ & $-208.4$ & $7.2$ \\
        \hline\hline
    \end{tabular}
    \label{tab:fan parameters}
\end{table}

\section{Results and discussion}
\label{sec:results}

\subsection{Axisymmetric cone cylinder shock}
\label{sec:results:cone}
First, we compare density fields obtained by conventional BOS algorithms to our physics-informed estimates. Section~\ref{sec:BOS:reconstruction} introduces three methods for axisymmetric BOS: Simpson's 1/3 Abel inversion, two-point Abel inversion, and unified BOS; the latter is conducted with a Sipkens kernel and Tikhonov regularization. Both Abel inversions require deflection data, which we obtain using a state-of-the-art wOFA algorithm \cite{Schmidt2021}. Conversely, the unified BOS and physics-informed algorithms operate directly on the unprocessed image difference data.\par

\begin{figure}[ht]
    \centering
    \includegraphics[width=5.75in]{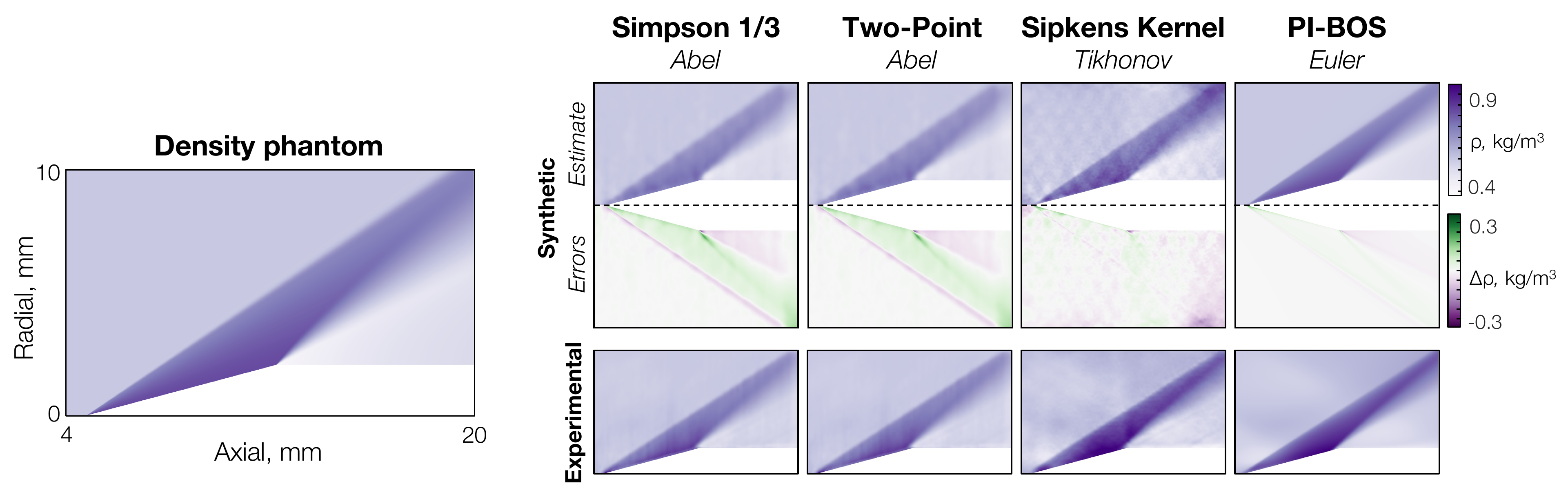}
    \caption{Phantom density field and reconstructions from noisy (single-shot) synthetic and experimental data. Errors are plotted for the synthetic case. Physics-informed BOS significantly increases the accuracy of reconstructions.}
    \label{fig:conventional reconstructions}
\end{figure}

Figure~\ref{fig:conventional reconstructions} depicts our Mach 2 cone cylinder density phantom as well as reconstructions computed with synthetic data (single-shot, noisy) and experimental data. The CFD simulation was designed to replicate the experimental conditions, so the phantom is expected to very nearly approximate true flow fields from the experiment. All BOS techniques considered in this paper produce a qualitatively accurate estimate of the density field. Uniform inflow compresses as it is deflected over the cone cylinder and expands as it flows past the shoulder. However, there is a clear difference in quality between these techniques. The explicit inverse Abel transforms underpredict compression, overpredict expansion, and exhibit a distinct error at the top-right corner that is associated with the Abel transform (namely, with the erroneous assumption of parallel rays). Results from these algorithms are substantially similar. Next, unified BOS with a Sipkens kernel produces a superior prediction in terms of magnitude, although artifacts from the sine wave background are clearly visible in the density field. These artifacts can be suppressed by increasing $\omega_\mathrm{Tik}$, but doing so damps the magnitude of the field and thereby amplifies reconstruction errors. Lastly, the physics-informed reconstruction almost perfectly recovers the density field, exhibiting crisp features, reminiscent of the ground truth density field, including a well-defined expansion fan.\par

\begin{figure}[ht]
    \centering
    \includegraphics[width=2.75in]{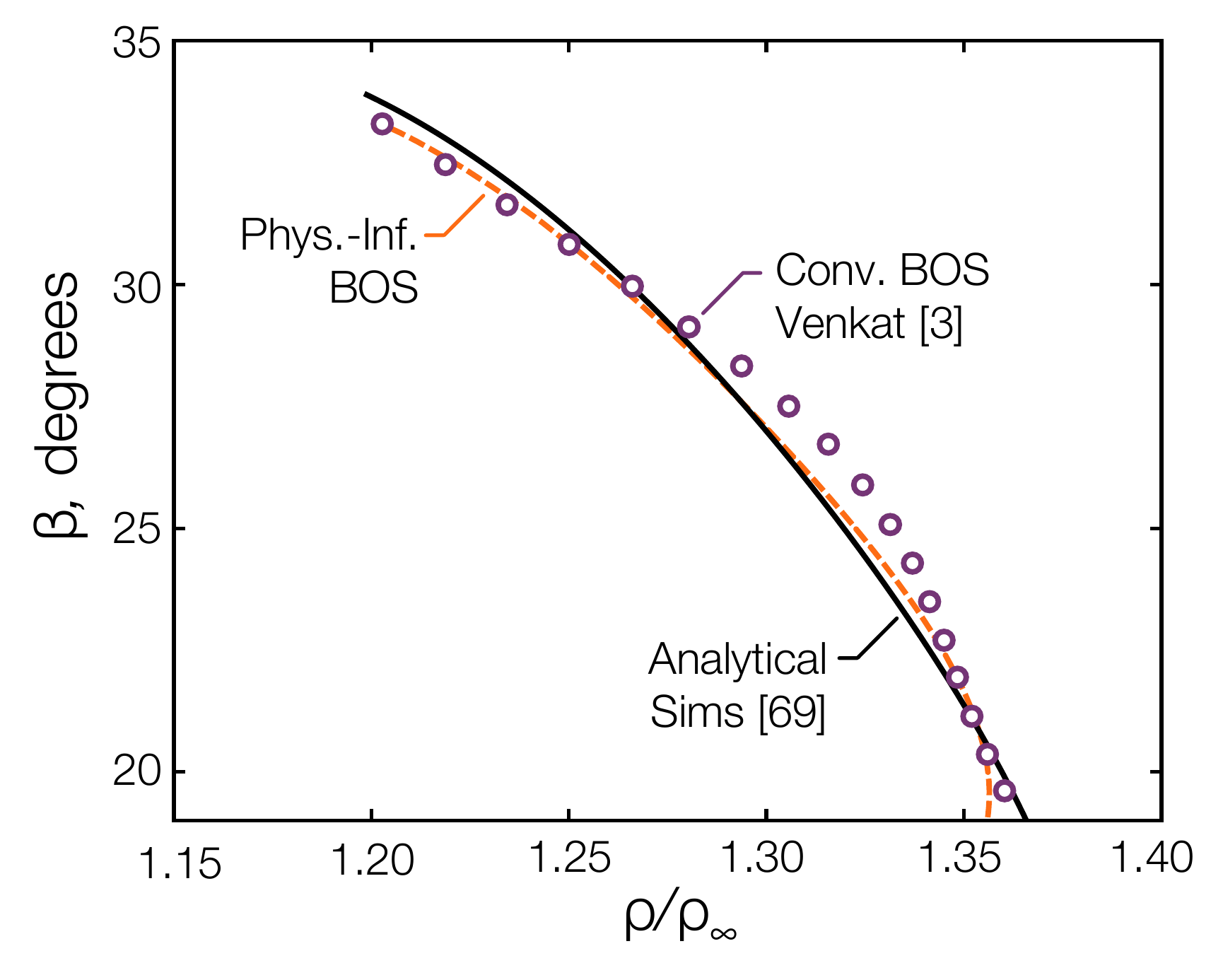}
    \caption{Density ratio behind the forward shock, 2~mm downstream of the cone apex, as a function of angular elevation with respect to the axis of symmetry. Results are compared to previous experimental \cite{Venkatakrishnan2004} and analytical cone table \cite{Sims1964} data.}
    \label{fig:density cut}
\end{figure}

To quantify relative performance, we calculate normalized root-mean-square errors (NRMSEs) for each reconstruction in percentage form. Both Simpson's 1/3 rule and two-point Abel inversion generate 8.34\% error and the Sipkens kernel with Tikhonov smoothing produces 6.12\% error for this phantom. By comparison, our physics-informed density field estimate has an error of only 3.75\%. While it is not possible to comprehensively baseline the accuracy of our experimental reconstruction, we can compare the resultant density field to predicted values from an analytical cone shock table \cite{Sims1964}. This comparison is shown in Fig.~\ref{fig:density cut}, alongside data from Venkatakrishnan and Meier \cite{Venkatakrishnan2004}. Those authors used a 5.1~MP camera (Sony DSC F-707) to record BOS images of the same cone shock structure, and they reconstructed the density field via filtered back projection.\footnote{Images from the 5.1~MP data set, recorded in 2003, are no longer available for processing.} The plot corresponds to a density cut taken 2~mm downstream of the cone apex. Our reconstruction neatly matches the analytical result, exhibiting a closer correspondence than the results from \cite{Venkatakrishnan2004}. Note that we omitted the 4$^\circ$ region immediately above the cone, which was severely affected by blur and calibration errors. Altogether, these results suggest that our physics-informed technique yields a qualitatively and quantitatively accurate estimate of the true field. Furthermore, to the best of our knowledge, this represents the first use of a PINN to reconstruct experimental measurements of a supersonic flow.\par

\begin{figure}[t]
    \centering
    \includegraphics[width=4.75in]{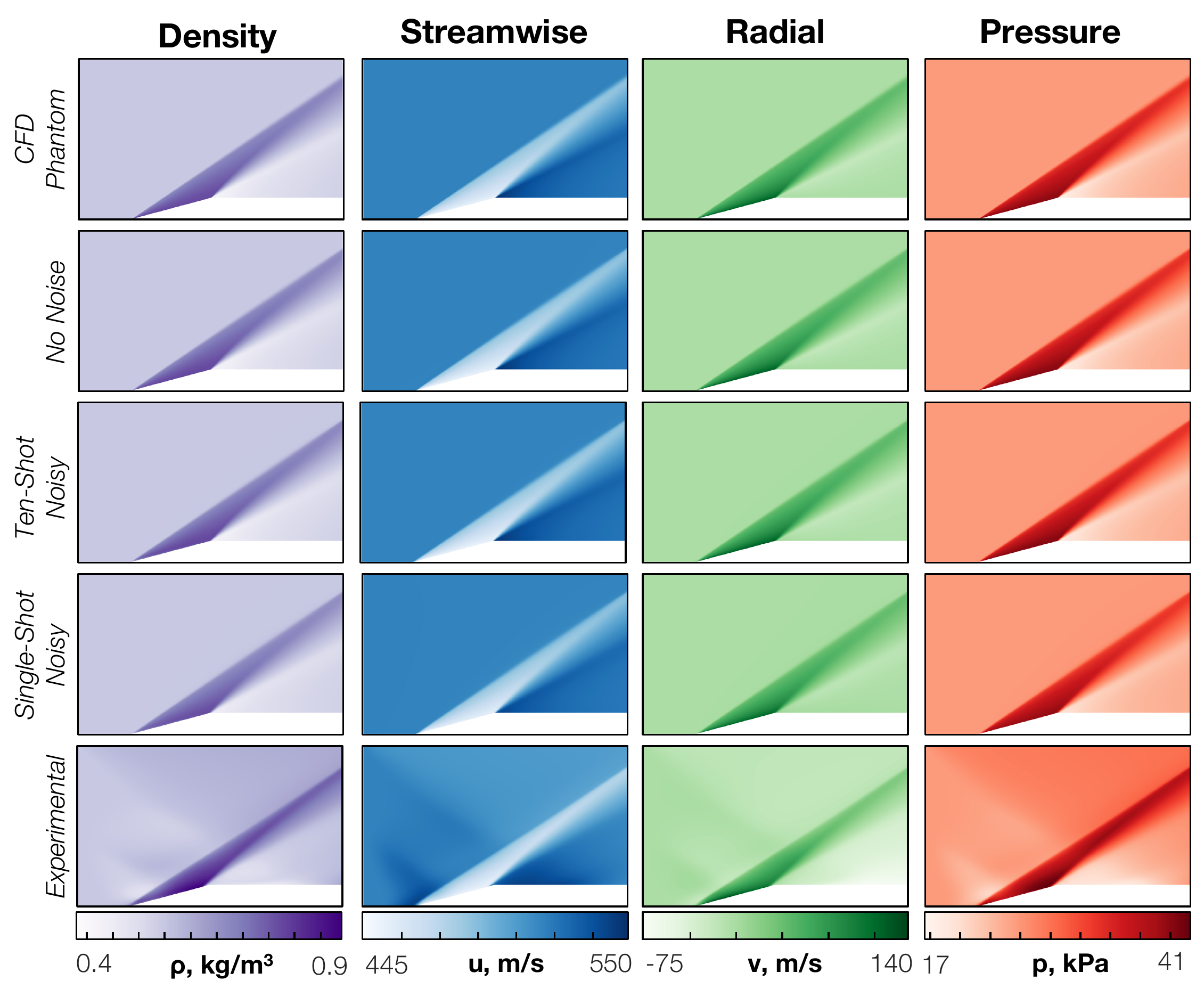}
    \caption{Reconstructions of all axisymmetric cone shock data sets. Synthetic data are reconstructed with increasing noise (clean, ten-shot average, and single-shot), followed by experimental data. All reconstructions (including the experimental case) bear a close resemblance to the CFD fields.}
    \label{fig:panel:cone}
\end{figure}

Not only does physics-informed BOS increase the accuracy of density field estimates, it also provides access to the latent fields. Fundamentally, the PINN outputs density, velocity, and total energy fields, which may be used to compute pressure or temperature, as desired. Velocity and pressure are directly relevant to the analysis of aerodynamic performance, so we present density, velocity, and pressure fields obtained from the cone cylinder measurements. These fields are estimated from synthetic and experimental data using physics-informed BOS. NRMSEs for all estimates computed with synthetic data are provided in Table~\ref{tab:errors}; the corresponding panel of phantoms and reconstructions can be found in Fig.~\ref{fig:panel:cone}.\par

\begin{table}[ht]
    \renewcommand{\arraystretch}{1.25}
    \caption{Percentage NRMSEs for Cone Shock Reconstructions from Best-Case Conventional and Physics-Informed BOS}
    \centering
    \begin{tabular}{c c c c c c}
        \hline\hline
        \multirow{2}{*}{Data} &
        \multicolumn{2}{c}{Density} &
        \multirow{2}{*}{$u$-velocity} &
        \multirow{2}{*}{$v$-velocity} &
        \multirow{2}{*}{Pressure}\\
        & \multicolumn{1}{c}{Conv.} & \multicolumn{1}{c}{PI-BOS} \\
        \hline
        Clean & \multirow{2}{*}{$4.81$} & $1.13$ & $0.36$ & $0.53$ & $1.39$\\
        Clean (no $\varepsilon_5$) &  & $1.22$ & $1.39$ & $0.74$ & $1.56$\\
        Averaged & \multirow{ 2}{*}{$4.98$} & $3.16$ & $0.47$ & $0.70$ & $1.86$\\
        Avg. (no $\varepsilon_5$) & & $3.24$ & $2.82$ & $1.30$ & $2.07$\\
        Single-shot & \multirow{ 2}{*}{$6.12$} & $3.75$ & $0.61$ & $1.13$ & $3.09$\\
        SS (no $\varepsilon_5$) & & $4.16$ & $2.09$ & $1.95$ & $3.42$\\
        \hline\hline
    \end{tabular}
    \label{tab:errors}
\end{table}

All fields are accurately reconstructed from synthetic measurements, with a maximum error of 3.75\%, 1.13\%, and 3.09\% in the density, velocity, and pressure estimates, respectively, when the full loss is employed. This high degree of accuracy corresponds to a clear visual resemblance in all cases. Indeed, differences between the phantoms and reconstructions of synthetic data are nearly imperceptible, regardless of the level of noise. In subsequent tests with \textit{additional} Gaussian errors, having a standard deviation up to 10\% of the intensity range, reconstruction errors for the ten-shot data remained below 5\%. However, the errors did increase with added noise (as expected), and visible artifacts are present in the experimental reconstruction due to significant noise in the image data and erratic background pattern gradients. Of particular note, all non-zero differences upstream of the compression shock in Fig.~\ref{fig:data panel} are erroneous, indicating the intensity and spatial extent of the noise. Nevertheless, our experimental results are consonant with our CFD results, which supports the use of physics-informed BOS. In addition, a Bayesian framework could be leveraged to counteract biased and correlated noise \cite{Molnar2022}. We also be note that the PINN contains fewer parameters than the CFD simulation, having 363,200 and 1,189,492 parameters, respectively, although the CFD solver has a runtime of approximately 10 min (as opposed to the PINN's several-hour training time).\par

\begin{figure}[ht]
    \centering
    \includegraphics[width=4.75in]{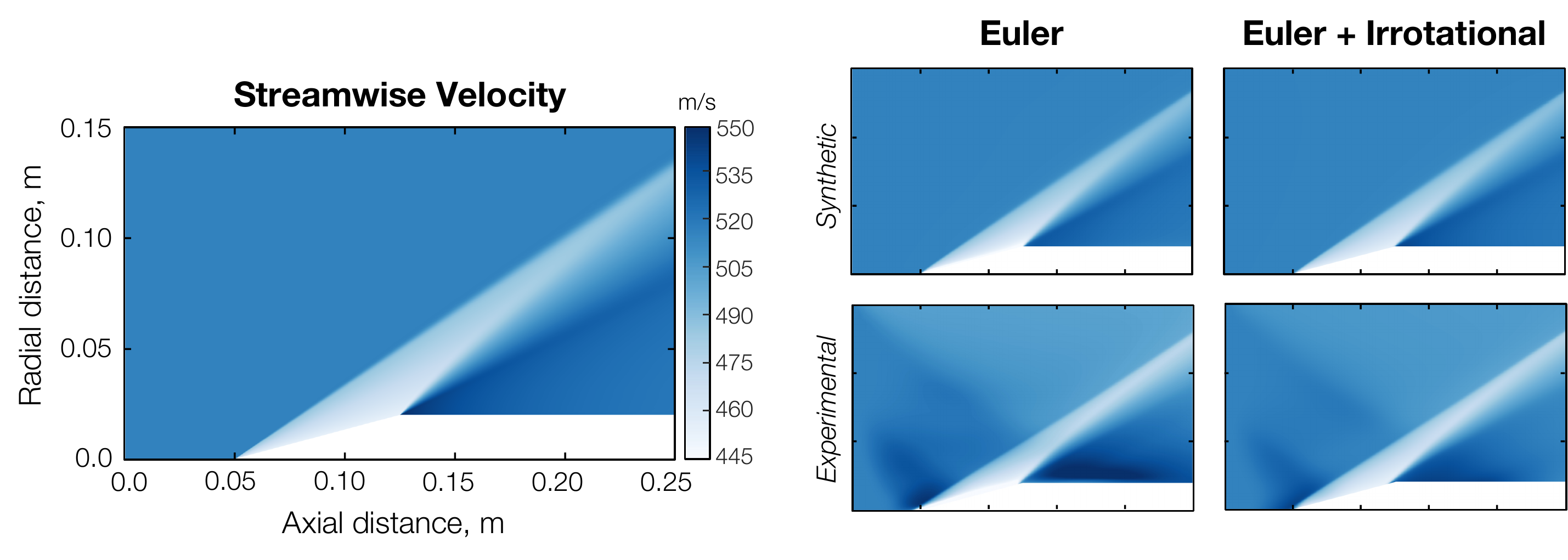}
    \caption{Streamwise velocity (left) and single-shot synthetic and experimental reconstructions (right). Incorporating an irrotationality equation provides an improvement that is enhanced in the presence of noise, as for the experimental images.}
    \label{fig:irrotational comparison}
\end{figure}

Table~\ref{tab:errors} includes two sets of errors for each image pair: one based on the whole physics loss (i.e., with the Euler and irrotationality equations, $\varepsilon_1$--$\varepsilon_4$ and $\varepsilon_5$) and one without $\varepsilon_5$. The resultant reconstructions are shown in Fig.~\ref{fig:irrotational comparison}. The utility of adding an irrotational flow residual to $\mathcal{L}_\mathrm{phys}$ is marginal for the synthetic cases. However, the benefit of including $\varepsilon_5$ increases with noise, per Table~\ref{tab:errors}. Consequently, there is a significant qualitative improvement in the experimental results when $\varepsilon_5$ is used, which manifests as a clear reduction of artifacts, with an acute improvement in the magnitude of the velocity field past the expansion fan. This result suggests that including additional terms, where appropriate, such as an entropy pair residual, could further improve the stability and accuracy of physics-informed BOS.\par

\subsection{Planar expansion fan}
\label{sec:results:fan}
\begin{figure}[ht]
    \centering
    \includegraphics[width=4.75in]{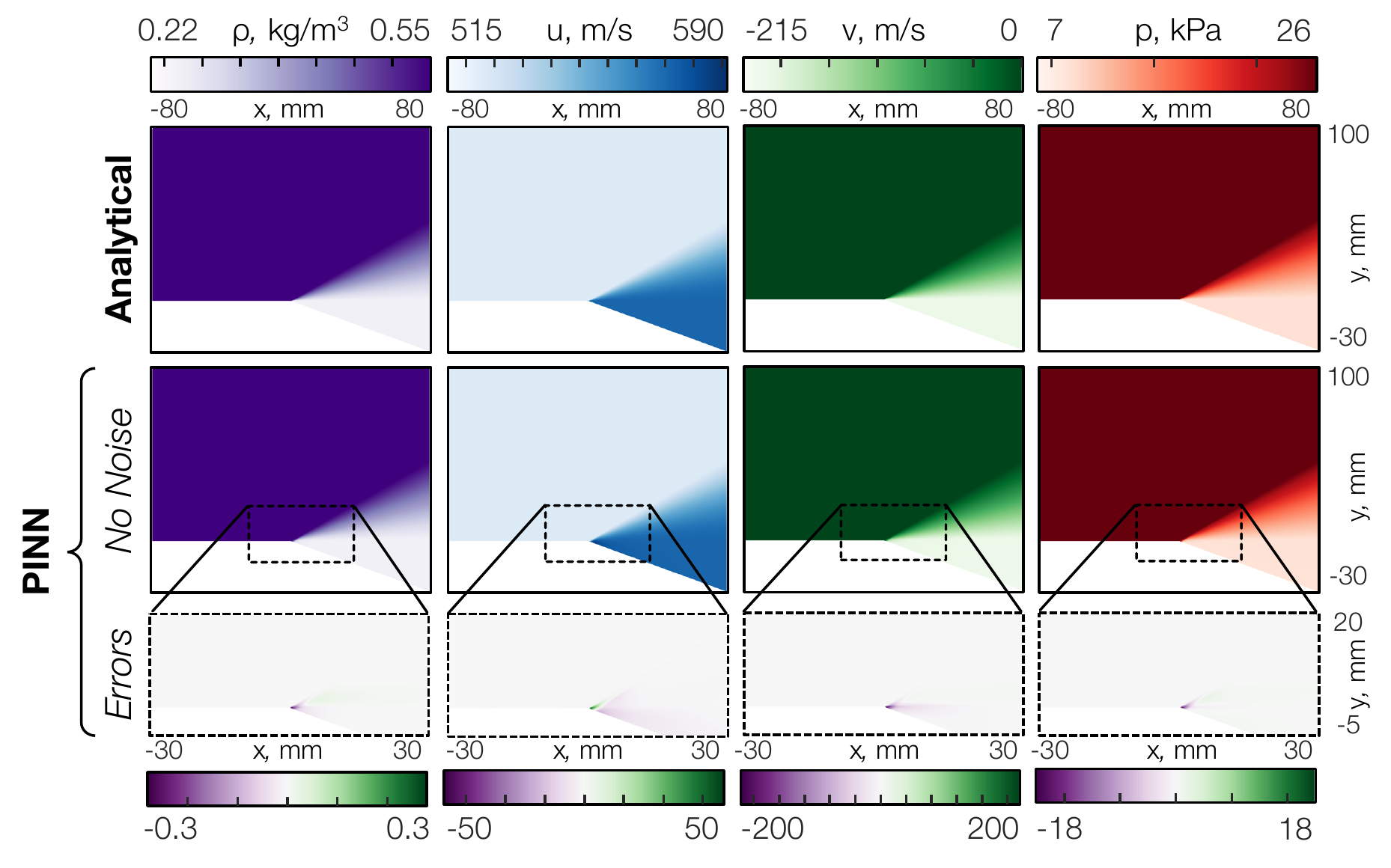}
    \caption{Reconstructions of a planar expansion fan from clean synthetic data. Errors are concentrated about the singularity.}
    \label{fig:panel:fan}
\end{figure}

Second, we present results for the Prandtl--Meyer expansion fan. A very similar set of fields was used by Jagtap and coworkers \cite{Jagtap2022}, allowing for a direct comparison between our PINN implementations. Figure~\ref{fig:panel:fan} depicts physics-informed BOS estimates of the fan's density, velocity, and pressure fields, all of which are recovered with high fidelity. The reconstructions exhibit minimal errors which are concentrated in the immediate vicinity of the singularity, located at the corner of the wedge. Including this region in the \textit{data} loss term can cause the solution to blow up, resulting in nonsensical fields. We tested this using a BOS data loss as well as the local density gradient loss term employed by Jagtap et al \cite{Jagtap2022}. The instability is present for both loss formulations. To avoid this issue, we omit a minute region surrounding the corner from our data loss. Excluding these gradients is justified for BOS, in any case, because they are too strong to satisfy the paraxial assumption, even for the narrow test section of interest, i.e., 3.75~cm. By contrast, physics residuals are computed throughout the whole domain.\par

\begin{figure}[ht]
    \centering
    \includegraphics[width=4.75in]{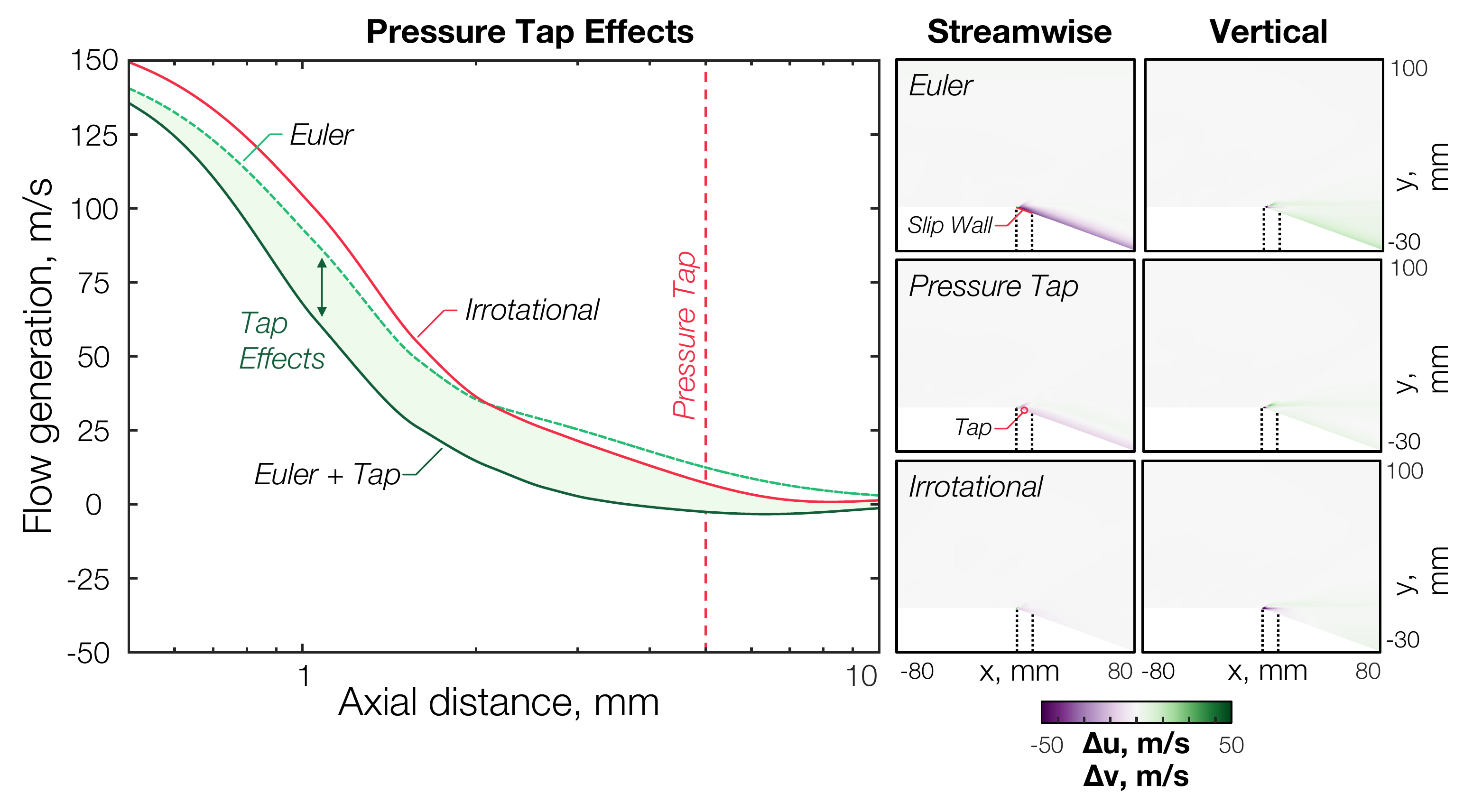}
    \caption{Flow generation along the first ten millimeters of the 80~mm ramp, plotted for distinct loss combinations (left). Axial and transverse velocity errors for each loss combination (right). Dashed lines in the error maps indicate the region over which flow generation is plotted.}
    \label{fig:flow generation}
\end{figure}

We also explored the stability of expansion fan reconstructions to varying measurement sources and physics loss components. Figure~\ref{fig:flow generation} depicts ``flow generation'' at the tip of the wedge and velocity error maps for different loss component scenarios. Positive flow generation indicates regions where the slip wall condition is not satisfied due to an erroneous inflow of momentum. We do not include a slip wall boundary condition in $\mathcal{L}_\mathrm{total}$, so this behavior is inferred from the BOS data, Euler equations, and inlet conditions, alone. However, there is a short region at the leading edge of the ramp over which we observe flow generation errors. These errors quickly diminish with distance normal to the ramp and down the ramp. Including an additional measurement towards the leading edge of the ramp, such as a pressure tap, significantly diminishes flow generation (note the $x$-axis is plotted in log scale).\par

While reconstructions of synthetic cone shock data are robust to the inclusion of $\varepsilon_5$ in $\mathcal{L}_\mathrm{phys}$, fan reconstructions are more sensitive, possibly due to a greater multiplicity of weak solutions to the Euler equations at the zero-viscosity limit. Streamwise velocity errors decrease from 1.15\% for a simple Euler loss to 0.61\% when the pressure tap is included and to 0.48\% for the Euler equations plus $\varepsilon_5$ (all other fields exhibit similar errors across these scenarios). In the first case, considering only the Euler equations, the PINN yields a significant over-prediction of streamwise velocity and under-prediction of the vertical component along the wedge. This is largely remedied by adding a single pressure tap 5~mm past the lip of the wedge, underlining the general principle that multi-modal measurements should be utilized whenever possible. Similar to Mao et al. \cite{Mao2020}, we find that the location of the pressure tap is an important consideration. Moving the tap much further down the wedge, e.g., to 75~mm, fully eradicates the gains to accuracy that we observe in the 5~mm case. This fade-out occurs because the flow generation problem is local and the tap is most useful in a region where image data is lacking (recall that we exclude BOS data near the singularity in this case). Unfortunately, while a single tap can reduce flow generation, it is insufficient to fully mitigate velocity field errors. Full-field constraints, such as the irrotationality equation, provide much more utility, as can be seen in the error maps in Fig.~\ref{fig:flow generation}.\par

In addition to these BOS-related tests, the expansion fan scenario serves as a convenient testbed for analyzing the performance of PINNs used for hyperbolic problems. For instance, we tested the dimensional and non-dimensional Euler equations in training and found that the former physics loss is far less stable.\footnote{Note that the dimensional and non-dimensional Euler equations have the same form.} This finding confirms the numerical tests of Haghighat et al. \cite{Haghighat2022}, who attributed the performance differential to the disparate magnitudes of the \textit{dimensional} Euler loss components, spanning roughly five orders of magnitude in our scenarios. An adaptive weighting scheme could potentially help to overcome this issue, but we previously observed that such schemes are unstable for noisy data \cite{Molnar2022}. Therefore, we recommend using the non-dimensional Euler equations, instead. Separately, we tested a suite of activation functions, including the swish, GELU, hyperbolic tangent, leaky ReLU, and sigmoid functions. To lacklustre effect, the performance of these functions was quite similar, although each choice exhibited a unique optimum of $\omega_\mathrm{meas}$, $\omega_\mathrm{phys}$, and $\omega_\mathrm{in}$, likely because some activation functions saturate (e.g., tanh and sigmoid) while others do not (e.g., swish and GELU).\par

\section{Conclusions and outlook} 
\label{sec:conclusions}
We present a data assimilation technique for background-oriented schlieren called physics-informed BOS. This approach combines a comprehensive measurement model, based on unified BOS, with the governing physical equations, in this case the Euler and irrotational equations, to infer steady density, velocity, and energy fields from a single pair of images. The method utilizes a physics-informed neural network to represent the flow; PINNs are flexible, easy-to-use tools for DA, but they had not previously been used to reconstruct supersonic flow from experimental data. Indeed, this work reports the first such use of a PINN to the best of our knowledge. We report accurate, multi-parameter reconstructions without the use of pressure data, artificial viscosity regularization, or an entropy condition. Several important conclusions can be drawn from this work.
\begin{enumerate}
    \item Physics-informed BOS with a PINN yields more accurate estimates of the density fields than conventional BOS algorithms. In all synthetic cases, differences between the phantoms and reconstructions are nearly imperceptible. Additionally, in the experimental case, there is good agreement with the CFD solution.
    
    \item Additional physics-based residuals, such as an irrotationality equation, can be included in the physics loss to improve the stability of training, accuracy of reconstructions, and resilience to noise. In much the same way, multi-modal measurement information can guide the PINN towards the correct, physical solution during training.
    
    \item Akin to conventional CFD techniques, singularities pose a significant challenge for PINNs and need to be addressed. In scenarios where a singularity may arise, physics constraints should be strengthened to stabilize the DA scheme.
    
    \item Tuning the relative weight of loss components is essential to the procedure, and the optimal weighting depends upon one's choice of activation functions.
\end{enumerate}\par

\appendix
\renewcommand{\thesection}{Appendix \Alph{section}}

\section{Bases and kernels for axisymmetric BOS tomography}
\label{app:bases and kernels}

\subsection{Bases}
\label{app:bases and kernels:bases}
Recalling Eq.~\eqref{equ:discrete field}, the radially-symmetric density field is approximated as a sum over the basis, $\Phi$,
\begin{equation}
    \rho\mathopen{}\left(x, r\right) \approx \sum_{j}^N \rho_j \,\varphi_j\mathopen{}\left(x, r\right),
    \label{equ:discrete field:radial}
\end{equation}
where $\varphi_j$ is the $j$th basis function. Typically, these functions are uniform, linear, or quadratic in BOS tomography. Following Sipkens et al. \cite{Sipkens2021}, we utilize a uniform discretization scheme in the axial direction and a linear scheme in the radial direction.\par

The uniform axial functions are given by
\begin{equation}
    \eta_j\mathopen{}\left(x\right) = H\mathopen{}\left(x - x_j\right) - H\mathopen{}\left(x - x_{j+1}\right),
    \label{equ:uniform basis}
\end{equation}
for a set of axial locations, $x_j$, where $H(\cdot)$ is the Heaviside function. Differentiating $\eta$ with respect to $x$ yields
\begin{equation}
    \frac{\partial \eta_j}{\partial x} = f_\updelta\mathopen{}\left(x - x_j\right) - f_\updelta\mathopen{}\left(x - x_{j+1}\right),
    \label{equ:uniform basis gradient}
\end{equation}
where $f_\updelta$ is the Dirac delta function. Our radial basis comprises piecewise linear functions that span circular annuli,
\begin{align}
    \phi_j\mathopen{}(r) =
        \begin{cases}
        \frac{r - r_{j-1}}{r_j - r_{j-1}} & r_{j-1} < r \leq r_j \\
        \frac{r - r_{j+1}}{r_j - r_{j+1}} & r_j < r \leq r_{j+1} \\
        0 & \text{otherwise}
        \end{cases}.
    \label{equ:linear basis}
\end{align}
Note that this expression must be modified at the first and final functions to exclude the $j-1$ and $j+1$ terms, as appropriate, assuming a uniform inner circle for the first basis function and a field that decays to zero at the outer radius. The radial gradient of $\phi_j$ is
\begin{align}
    \frac{\partial \phi_j}{\partial r} = 
        \begin{cases}
        \frac{1}{r_j - r_{j-1}} & r_{j-1} < r \leq r_j \\
        \frac{1}{r_j - r_{j+1}} & r_j < r \leq r_{j+1} \\
        0 & \text{otherwise}
        \end{cases}.
        \label{equ:linear basis gradient}
\end{align}
Lastly, we combine these bases to obtain an axisymmetric linear--uniform basis,
\begin{equation}
    \rho\mathopen{}\left(x, r\right) \approx \sum_j^N \rho_j \,\underbrace{\eta\mathopen{}\left(x\right) \phi\mathopen{}\left(r\right)}_{\varphi_j\mathopen{}\left(x, r\right)},
    \label{equ:2D basis}
\end{equation}
where $\varphi_j$ is the resultant 2D basis function. The second-order Tikhonov matrix for this basis has elements
\begin{equation}
     L_{i,j} =
     \begin{cases}
        -4, & i = j\\
        1, & \text{$i$ axially adjacent to $j$} \\
        \left(2r_i \pm h \right)/\left(2r_i\right), & \text{$i$ radially adjacent to $j$}  \\
        0, & \text{otherwise}
    \end{cases},
    \label{equ: Tikhonov}
\end{equation}
where $\pm$ corresponds to $+$ and $-$ for outer and inner radial neighbors, respectively, and each element is scaled by the grid spacing, $h^{-2}$, assuming equal spacing in $x$ and $r$, i.e., $h = \Delta x = \Delta r$. This definition is modified to enforce the free stream refractive index at the outer radius. Effectively, this penalty utilizes the discrete cylindrical Laplacian of $\rho$ to promote reconstructions that are both axially- and radially-smooth.\par

\subsection{Kernels}
\label{app:bases and kernels:kernels}
Three direct, conventional kernels are utilized to baseline the PINN approach. In this context, ``direct'' means that the kernel relates the deflection data to a density field (Simpson's 1/3 rule and the two-point method) or vice versa (the Sipkens kernal).\footnote{We incorporate the Sipkens kernel into a unified BOS operator, which relates the density field directly to unprocessed image differences, although unified BOS is not inherent to that scheme.} ``Indirect'' refers to methods that require an explicit Poisson solver.\par

Simpson's rule arises from a traditional approach to numerical integration in which the area under a curve is approximated using the area under parabolic curves. Similarly, the two-point method utilizes piecewise linear interpolation with a quadratic expansion about the singularity. For a smooth target function with sufficient resolution, these approximations are highly similar. Both techniques yield a kernel, $\mathbf{K}$, that relates $\boldsymbol\updelta_\mathrm{y}$ to a radial density field, $\boldsymbol\uprho$. Separately, the Sipkens kernel, derived in \cite{Sipkens2021}, is a discrete forward operator that relaxes the assumption of parallel rays. Sipkens et al. \cite{Sipkens2021} developed the transform for a variety of bases; we utilize the linear--uniform basis introduced above. The result is deflection operators, $\mathbf{D}_\mathrm{x}$ and $\mathbf{D}_\mathrm{y}$, that are used to construct a unified BOS measurement model.\par

\subsubsection{Simpson's 1/3 rule}
\label{app:bases and kernels:kernels:Simpson}
Simpson's 1/3 rule involves piecewise curve fits to the diverging integral. The kernel has elements
\begin{subequations}
    \begin{align}
        K_{i,j} &=
        \begin{cases}
            0, & i > j \\
            J_{i,j}/2, & j = N + 1\\
            J_{i,j}, & i < j \ \mathrm{and} \ j \neq N+1\\
        \end{cases},
        \label{equ:Simpson}
        \intertext{where}
        J_{i,j} &= -\frac{1}{3\pi} \frac{3 + (-1)^{j-i+1}}{ \sqrt{\left(j-1 \right)^2 - \left(i-1 \right)^2}}
        \label{equ:Simpson:inner}
    \end{align}
\end{subequations}
and $N$ is the number of equally spaced intervals. It should be noted that the integrand diverges at the lower limit of integration. Therefore, the singularity is extrapolated from the remaining coefficients. This is expressed in the kernel as
\begin{equation}
    K_{i,i} =
    \begin{cases}
        K_{i, i+1}, & i \leq N \\
        0, & i = N + 1
    \end{cases}.
    \label{equ:Simpson:extrapolation}
\end{equation}
One key difference between Simpson's rule and the two-point method is that the former requires an even number of intervals due to the curve fitting technique, which uses three points. Nevertheless, Simpson's rule exhibits greater accuracy and faster convergence than integration by many other such rules (e.g., trapezoidal).\par

\subsubsection{Two-point method}
\label{app:bases and kernels:kernels:two-point}
When using the two-point method, the integral is discretized between neighboring radii and the deflection is presumed constant between those two rings (hence the name). The kernel is formed using pairs of projections, resulting in elements
\begin{equation}
    K_{i,j} =
    \begin{cases}
        0, & i > j \\
        J_{i,j}, & i = j \\
        J_{i,j} - J_{i, j-1}, & i < j
    \end{cases},
    \label{equ:two-point}
\end{equation}
where
\begin{equation}
    J_{i,j} =
    \begin{cases}
        0, & i > j \\
        2/\pi, & i = j = 0 \ .\\
        \frac{1}{\pi} \log\mathopen{}\left[ \frac{\sqrt{\left( j+1 \right)^2 - i^2} + j + 1}{\sqrt{\left(j-1 \right)^2 - i^2} + j} \right] & i \leq j
    \end{cases}
    \label{equ:two-point:inner}
\end{equation}
This formulation has no smoothing. The lack of smoothing can be an advantage over other reconstruction kernels given noise-free data, but it comes the a cost of noise amplification. The resulting deconvolution kernel is an upper triangular matrix, and the divergence of the integral is handled via a quadratic expansion of the projections at the singularity \cite{Cormack1982}.\par

\subsubsection{Sipkens deflectometry}
\label{app:bases and kernels:kernels:Sipkens}
The Sipkens kernel is a forward operator that can model deflections along arbitrary, non-parallel rays. This formulation can simultaneously operate on deflection data above and below the axis of symmetry, unlike the explicit inverse kernels described above, which improves the stability and resolution of reconstructions. Moreover, the forward kernel can be incorporated into any inverse solver, such as our physics-informed BOS technique. Here, we recall the final form of the linear--uniform Sipkens deflectometry operator using the present notation; a full derivation can be found in the supplementary material of \cite{Sipkens2021}.\par

To start, the kernel requires a 3D description of each ray. Consider the $i$th ray as it passes by the $z$-axis at the point $(x_i, y_i, 0)$ with a slope of $m_{\mathrm{x},i}$ in the $x$-$z$ plane and $m_{\mathrm{y},i}$ in the $y$-$z$ plane. For ease of notation, we introduce a function that corresponds to the radius of the $i$th ray at this crossing point,
\begin{equation}
    r_{\mathrm{ray},i}\mathopen{}\left(x\right) = \frac{1}{m_{\mathrm{x},i}} \sqrt{\left(x - x_i \right)^2 + \left[m_{\mathrm{y},i} \left(x - x_i \right) + y_i \right]^2}.
    \label{equ:radius function}
\end{equation}
Each ray is divided into two, with one on either side of $z = 0$, i.e., a ray approaching the central axis and a ray departing from it. Consequently, the vertical deflectometry operator is split into two parts,
\begin{equation}
    D_{\mathrm{y},i,j} = D_{\mathrm{y},i,j}^+ + D_{\mathrm{y},i,j}^-.
    \label{equ:bifurcation}
\end{equation}
The ``approaching'' and ``departing'' elements have a very similar expression, which we present in a consolidated form,
\begin{align}
    \begin{split}
        D_{\mathrm{y},i,j}^\pm &= \frac{\sqrt{1 + m_{\mathrm{x},i}^2}}{1 + m_{\mathrm{y},i}^2}
        \sum_j^N \left( \vphantom{\frac{0_j}{0_j}} y_{0,i} \left\{
        a_\mathrm{int}\mathopen{}\left[r_{j-1}, r_j, r_2\mathopen{}\left(j\right) \right] -
        a_\mathrm{int}\mathopen{}\left[r_{j-1}, r_j, r_1\mathopen{}\left(j\right)\right] +
        a_\mathrm{int}\mathopen{}\left[r_{j+1}, r_j, r_3\mathopen{}\left(j\right) \right]
        \right. \right. \\ & \qquad \left. \left. -
        a_\mathrm{int}\mathopen{}\left[r_{j+1}, r_j, r_2\mathopen{}\left(j\right) \right] \right\}
        \pm m_{\mathrm{y},i} \left[ \frac{r_\mathrm{b}\mathopen{}\left(j\right) - r_\mathrm{b}\mathopen{}\left(j-1\right)}{r_j - r_{j-1}} - \frac{r_\mathrm{b}\mathopen{}\left(j\right) - r_\mathrm{b}\mathopen{}\left(j+1\right)}{r_j - r_{j+1}} \right] \right),
    \end{split}
    \label{equ:radial operator}
\end{align}
where $\pm$ becomes $+$ for $D_{\mathrm{y},i,j}^+$ and $-$ for $D_{\mathrm{y},i,j}^-$,
\begin{equation}
    a_\mathrm{int}\mathopen{}\left(r_1,r_2,r_3\right) = \frac{1}{r_2 - r_1} \,\mathrm{log}\mathopen{}\left\{ \mathrm{abs}\mathopen{}\left[r_3 + \left(r_3^2 - \frac{y_{0,i}^2}{1 + m_{\mathrm{y},i}^2} \right)^{1/2}\right] \right\},
    \label{equ:radial operator:a function}
\end{equation}
and $r_\mathrm{b}$ modifies the $j$th radius to incorporate the bounds of the Heaviside function. See \cite{Sipkens2021} for a discussion of the $r_1$, $r_2$, and $r_3$ functions of $j$ in Eq.~\eqref{equ:radial operator}.\par

The axial deflection operator is simply
\begin{equation}
    D_{\mathrm{x},i,j} = \mathrm{abs}\mathopen{}\left\{ \phi_j\mathopen{}\left[ r_{\mathrm{ray},i}\mathopen{}\left(x_j\right) \right] - \phi_j\mathopen{}\left[ r_{\mathrm{ray},i}\mathopen{}\left(x_{j+1}\right) \right] \right\}.
    \label{equ:axial operator}
\end{equation}
Since the axial basis is piecewise uniform, $\mathbf{D}_\mathrm{x}$ only has non-zero elements for rays that cross an axial boundary. While this representation is sub-optimal, it is sufficient for the present demonstration. A bi-linear basis will be derived for future use. Equations~\eqref{equ:radial operator} and \eqref{equ:axial operator} are employed to populate $\mathbf{D}_\mathrm{y}$ and $\mathbf{D}_\mathrm{x}$, which are themselves required to construct the unified BOS measurement model in Eq.~\eqref{equ:unified BOS}.\par

\section{PINNs applied to hyperbolic equations}
\label{app:physics-informed BOS hyperbolic}
Here, we briefly review two key developments in the area of physics-informed neural networks applied to hyperbolic equations. This topic directly pertains to PINNs discussed in this work because the Euler equations are hyperbolic.\par

\subsection{Artificial viscosity}
\label{app:physics-informed BOS hyperbolic:viscosity}
``Artificial viscosity'' was initially introduced into Euler solvers to accommodate discontinuous flow fields by spreading shocks across a finite region that could be resolved by the grid \cite{Neumann1950, Lax1959}. Artificial viscosity entails the deliberate addition of dissipation into the governing equations, often based on the size of the grid, such that the equations can be solved across the shock without introducing excessive errors.\par

Several researchers have applied this concept to PINNs to provide stability in the context of a hyperbolic physics loss. Fuks et al. \cite{Fuks2020} were the first to do this for two-phase transport in porous media, i.e., adding viscous damping to a Buckley--Leverett transport model. Coutinho and coworkers \cite{Coutinho2022} adopted a similar approach and considered the inviscid Burgers' equation in addition to a Buckley--Leverett scenario; Coutinho also introduced a variety of adaptive methods to automatically tune the dissipation term. More recently, Patel et al. \cite{Patel2022} proposed the use of a control volume physics loss with artificial viscosity to solve a 1D shock governed by Burgers' equation, a 1D Sod shock problem, and generic 1D Buckley--Leverett, Euler, and Leblanc problems. Collectively, these papers demonstrate that including artificial viscosity can reduce oscillations about a discontinuity and stabilize optimization of the network.\par

Most hyperbolic solvers with artificial viscosity require careful selection of the viscous term, which may vary across space and time. Excessive damping is known to yield stable yet invalid solutions \cite{Anderson1990}. This was noted by Liu et. al. \cite{Liu2022}, leading to their usage of a gradient-dependent $\omega_\mathrm{phys}$ parameter, which diminished in highly compressible regions, instead of an artificial viscosity scheme. Further, Wang et al.'s \cite{Wang2021} analysis of PINNs demonstrates that higher-order gradients dominate the cumulative gradient vector used to update the network parameters. This result suggests that a non-physical diffusion term can obfuscate progress towards the correct, physical solution. Since BOS measurements provide relatively direct access to the real density field, we do not employ artificial viscosity for shock capturing.\par

\subsection{Entropy pair regularization}
\label{app:physics-informed BOS hyperbolic:entropy}
Many hyperbolic systems of equations, including the compressible Euler equations, admit multiple ``weak'' solutions. The physical solution is often selected by applying a constraint like an entropy condition, which is valid for inviscid flows that exhibit one or more discontinuities. This condition leads to an inequality that relates mathematical entropy to the divergence of an entropy-flux pair \cite{Lellis2010}; the inequality is satisfied by desirable ``viscosity solutions'' to the governing equations.\par

Two groups have added an entropy loss to a PINN to learn shock-laden flow. Patel et al. \cite{Patel2022} first implemented this technique, using a complete equation of state that bounds the Euler equations by a specific entropy relation. Separately, Jagtap et al. \cite{Jagtap2022} used a flux pair coupled with a polytropic equation of state. Patel's formulation resulted in three loss components whereas Jagtap's method culminated in a scalar loss. Both groups reported that entropy pair regularization improved learning for hyperbolic problems compared to a vanilla PINN. However, it should be noted that a unique entropy pair is not guaranteed for a given set of hyperbolic conservation laws \cite{Godlewski2013}. Therefore, we chose to forego an entropy pair loss, which proved to be unnecessary for our test cases.\par

\section{Nonlinear Ray Tracing}
\label{app:ray tracing}
Synthetic data are generated using nonlinear ray tracing within the variable index field. This is accomplished to high accuracy via the fourth-order Runge--Kutta ray tracing algorithm of Sharma et al. \cite{Sharma1982}. First, Eq.~\eqref{equ:ray equation} is split into coupled ordinary differential equations. Next, each ray is defined by a starting position, $\mathbf{x}_0 = [x, y, z]^\mathrm{T}$, and a refractive index-scaled direction,
\begin{align}
    \mathbf{v}_0 = n\mathopen{}\left(\mathbf{x}\right) \frac{\mathrm{d}\mathbf{x}}{\mathrm{d}s}
\end{align} 
where $\mathrm{d}\mathbf{x}/\mathrm{d}s$ is the initial trajectory of the ray. Next, consider the vector valued function
\begin{equation}
    \mathbf{f}\mathopen{}\left(\mathbf{x}\right) = \Delta s \,n\mathopen{}\left(\mathbf{x}\right)
    \boldsymbol\nabla n\mathopen{}\left(\mathbf{x}\right) = \tfrac{1}{2} \,\Delta s
    \,\boldsymbol\nabla n^2\mathopen{}\left(\mathbf{x}\right).
\end{equation}
We evaluate $\mathbf{f}$ at three points along the ray,
\begin{equation}
    \begin{split}
        \mathbf{f}_\mathrm{A} &= \mathbf{f}\mathopen{}\left(\mathbf{x}_i\right) \\
        \mathbf{f}_\mathrm{B} &= \mathbf{f}\mathopen{}\left[\mathbf{x}_i + \Delta s \left(\tfrac{1}{2}\mathbf{v}_i + \tfrac{1}{8}\mathbf{f}_\mathrm{A}\right) \right], \\
        \mathbf{f}_\mathrm{C} &= \mathbf{f}\mathopen{}\left[\mathbf{x}_i + \Delta s \left(\mathbf{v}_i + \tfrac{1}{2}\mathbf{f}_\mathrm{B}\right) \right] \\
    \end{split}
\end{equation}
where the subscript $i$ denotes the current index along the path of the ray. The ray's position and direction are updated,
\begin{equation}
    \begin{split}
        \mathbf{x}_{i+1} &= \mathbf{x}_i + \Delta s \left[\mathbf{v}_i + \tfrac{1}{6} \left(\mathbf{f}_\mathrm{A} + 2\mathbf{f}_\mathrm{B}\right) \right] \\
        \mathbf{v}_{i+1} &= \mathbf{v}_i + \tfrac{1}{6} \left(\mathbf{f}_\mathrm{A} + 4\mathbf{f}_\mathrm{B} + \mathbf{f}_\mathrm{C}\right)
    \end{split},
\end{equation}
and the procedure is repeated until all rays exit the computational domain.\par

\section{Generation of flow phantoms}
\label{app:phantoms}
This appendix provides additional details about our phantoms.\par

\subsection{Axisymmetric cone cylinder}
\label{app:phantoms:cone cylinder}
We simulated the cone cylinder flow using SU2~7.3.0, which is a finite volume solver; convective fluxes are handled with a second-order upwind scheme in space that is TV diminishing when paired with a limiter \cite{Roe1981}. The Venkatakrishnan slope limiter is utilized to combat the oscillatory behavior of higher-order upwind schemes by enforcing a monotonicity condition \cite{Venkatakrishnan1995}. Additionally, gradients are computed using a weighted least squares algorithm.  We use a 2D axisymmetric unstructured grid that comprises 297,373 triangular cells, with moderate refinement towards the anticipated shock locations and tripping points. Slip-wall conditions, based on the inviscid flow assumption, are implemented along the cone cylinder and top of the domain, which represents the physical boundary of the wind tunnel. \par

\begin{figure}[ht]
    \centering
    \includegraphics[width=4.5in]{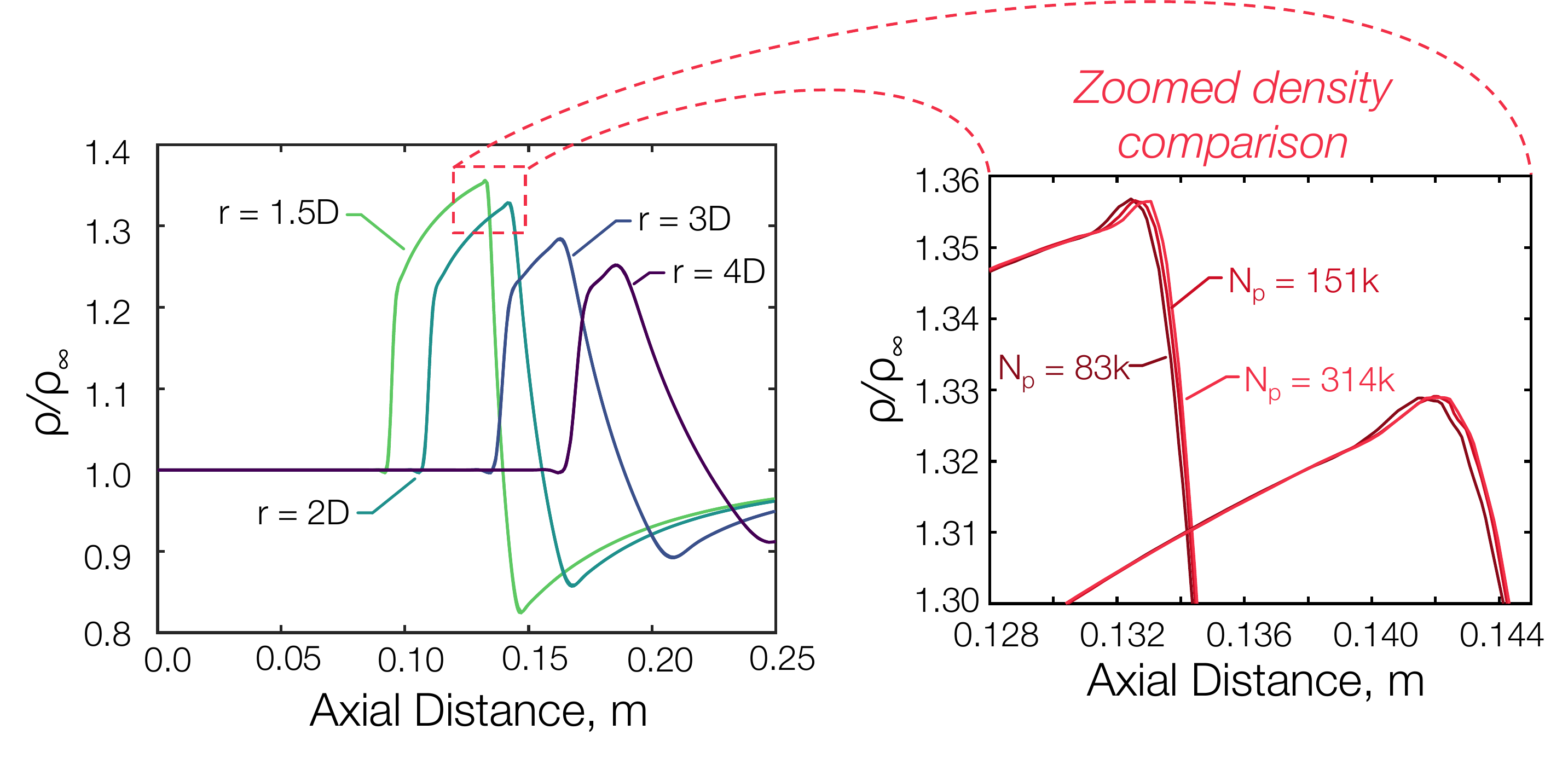}
    \caption{Axial cuts of the simulated cone shock density field for three computational grids (left). Cuts are shown at selected radii, and a zoomed comparison is presented (right). The CFD results exhibit minimal discrepancies, even across the shocks.}
    \label{fig:cfd convergence}
\end{figure}

A grid convergence study was performed to ensure that the solution was independent of the grid. Computations were performed with cells of size $1.25 \times 10^{-4}$~m, $6.25 \times 10^{-5}$~m, and $3.125 \times 10^{-5}$~m. Results from this study are shown in Fig.~\ref{fig:cfd convergence}. Radial slices of the non-dimensionalized density field are plotted, with a zoomed view provided to illustrate the close agreement between simulations. Notably, the density, velocity, and energy errors all asymptote, indicating that the shocks are well captured by all three schemes. Errors resulting from the computational grid are thus expected to be marginal. Throughout the paper, we use results computed with the intermediate grid, with $6.25 \times 10^{-5}$~m cells, to generate synthetic data.\par

\subsection{Planar expansion fan}
\label{app:phantoms:PM fan}
The planar expansion fan serves as a useful phantom to evaluate the BOS reconstruction method developed here, given the analytical nature. Specifically, this solution arises through a geometric analysis of a sequence of Mach waves, resulting in the governing differential equation for Prandtl--Meyer flow,
\begin{equation}
    \mathrm{d}\theta = \sqrt{M^2 - 1} \,\frac{\mathrm{d}V}{V},
    \label{equ:Prandtl-Meyer:differential}
\end{equation}
where $\theta$ is the turning angle towards the ramp and $V$ is the local flow speed. Equation~\eqref{equ:Prandtl-Meyer:differential} is expressed in terms of $\mathrm{d}M/M$ and integrated to the ramp to obtain the outflow Mach number,
\begin{equation}
    \int_{\theta_1}^{\theta_2} \mathrm{d}\theta= \int_{M_1}^{M_2} \frac{\sqrt{M^2 - 1}}{1 + \frac{\gamma - 1}{2}M^2} \frac{\mathrm{d}M}{M}.
    \label{equ:Prandtl-Meyer:integral}
\end{equation}
The indefinite solution to the right side of Eq.~\eqref{equ:Prandtl-Meyer:integral}, called the Prandtl--Meyer function, is
\begin{subequations}
    \begin{align}
        \nu\mathopen{}\left(M\right) &= \sqrt{\frac{\gamma + 1}{\gamma - 1}} \,\mathrm{tan}^{-1}\mathopen{}\left[\sqrt{\frac{\gamma - 1}{\gamma + 1} \left( M^2 - 1 \right)} \,\right] - \mathrm{tan}^{-1}\mathopen{}\left(\sqrt{M^2 - 1}\right), \label{equ:Prandtl-Meyer function:closed form}
        \intertext{which can be used to solve Eq.~\eqref{equ:Prandtl-Meyer:integral} across a differential wedge,}
        \theta - \theta_1 &= \nu\mathopen{}\left[ M\mathopen{}\left(\theta\right) \right] - \nu\mathopen{}\left( M_1 \right). \label{equ:Prandtl-Meyer function:differential form}
    \end{align}
    \label{equ:Prandtl-Meyer function}%
\end{subequations}
For a horizontal wall leading up to the ramp, $\theta_1$ is zero and the Mach number is computed as a function of the turning angle. This may be done by solving Eq.~\eqref{equ:Prandtl-Meyer function:differential form} via Newton's method,
\begin{equation}
    M^{(k+1)}\mathopen{}\left(\theta\right) = \frac{\theta + \nu\mathopen{}\left(M_1\right) - \nu\mathopen{}\left[M^{(k)}\mathopen{}\left(\theta\right)\right]} {\nu^\prime\mathopen{}\left[M^{(k)}\mathopen{}\left(\theta\right)\right]} + M^{(k)}\mathopen{}\left(\theta\right),
    \label{equ:Newton}
\end{equation}
where
\begin{equation}
    \nu^\prime\mathopen{}\left(M\right) = \frac{\sqrt{M^2 - 1}}{M \left( 1 + \frac{\gamma - 1}{2} M^2\right)}
    \label{equ:Newton:Prandtl-Meyer function:derivative}
\end{equation}
is the analytical derivative of the Prandtl--Meyer function. Once the Mach field has been determined, the other field variables are calculated using the isentropic relations,
\begin{subequations}
    \begin{align}
        T\mathopen{}\left(\theta\right) &= T_1 \left[\frac{1 + \frac{\gamma - 1}{2}M_1^2}{1 + \frac{\gamma - 1}{2} \,M\mathopen{}\left(\theta\right)^2} \right], \\
        p\mathopen{}\left(\theta\right) &= p_1 \left[ \frac{T\mathopen{}\left(\theta\right)}{T_1}\right]^\frac{\gamma}{\gamma-1}, \quad\text{and}\\
        \rho\mathopen{}\left(\theta\right) &=  \frac{p\mathopen{}\left(\theta\right)}{R_\mathrm{gas} \,T\mathopen{}\left(\theta\right)},
    \end{align}
\end{subequations}
where $R_\mathrm{gas} = 287$~J/kg\,K is the gas constant for air.\par

Although Mach numbers in the fan are calculated as a function of $\theta$, each Mach line is oriented at the corresponding Mach angle,
\begin{equation}
    \mu\mathopen{}\left(\theta\right) = \mathrm{sin}^{-1}\mathopen{}\left[ M\mathopen{}\left(\theta\right)^{-1} \right],
    \label{equ:Mach angle}
\end{equation}
as drawn in Fig.~\ref{fig:expansion fan}. The local flow speed is simply
\begin{equation}
    V = M\,a,
\end{equation}
where
\begin{equation}
    a = \sqrt{\gamma \,R_\mathrm{gas} \,T}
\end{equation}
is the speed of sound. Finally, the flow accelerates tangent to Mach lines throughout the fan such that the streamlines are parallel to the wall up to the forward Mach line, normal to the turning angle throughout the fan, and parallel to the wedge thereafter,
\begin{subequations}
    \begin{align}
        u &= V\,\cos\mathopen{}\left(\theta\right) \quad\text{and} \label{equ:fan velocity:u} \\
        v &= -V\,\sin\mathopen{}\left(\theta\right). \label{equ:fan velocity:v}
    \end{align}
    \label{equ:fan velocity}
\end{subequations}\par

\section*{Declarations}
\subsection*{Ethical Approval}
Not applicable.\par

\subsection*{Competing interests}
None.\par

\subsection*{Authors' contributions}
All authors made a significant contribution.\par

\subsection*{Funding}
This material is based upon work supported by the National Science Foundation under Grant No. 2227763, the Erlangen Graduate School in Advanced Optical Technologies (SAOT) at the Friedrich-Alexander-Universit{\"a}t Erlangen-N{\"u}rnberg, and a Department of Defense National Defense Science and Engineering Graduate Fellowship.\par

\subsection*{Availability of data and materials}
Data will be made available upon reasonable request.\par

\bibliographystyle{osajnl2} 

\begin{thebibliography}{10}
\newcommand{\enquote}[1]{``#1''}

\bibitem{Dolvin2008}
D.~Dolvin, \enquote{Hypersonic international flight research and
  experimentation {(HIFiRE)} fundamental science and technology development
  strategy,} in \enquote{15th AIAA International Space Planes and Hypersonic
  Systems and Technologies Conference,}  (2008), p. 2581.

\bibitem{Raffel2015}
M.~Raffel, \enquote{Background-oriented schlieren ({BOS}) techniques,} Exp.
  Fluids \textbf{56}, 1--17 (2015).

\bibitem{Venkatakrishnan2004}
L.~Venkatakrishnan and G.~Meier, \enquote{Density measurements using the
  background oriented schlieren technique,} Exp. Fluids \textbf{37}, 237--247
  (2004).

\bibitem{Sommersel2008}
O.~Sommersel, D.~Bjerketvedt, S.~Christensen, O.~Krest, and K.~Vaagsaether,
  \enquote{Application of background oriented schlieren for quantitative
  measurements of shock waves from explosions,} Shock Waves \textbf{18},
  291--297 (2008).

\bibitem{Yamagishi2021}
M.~Yamagishi, Y.~Yahagi, M.~Ota, Y.~Hirose, S.~Udagawa, T.~Inage, S.~Kubota,
  K.~Fujita, K.~Ohtani, and H.~Nagai, \enquote{Quantitative density measurement
  of wake region behind reentry capsule ({Improvements} in accuracy of {3D}
  reconstruction by evaluating the view-angle of measurement system),} J. Fluid
  Sci. Technol. \textbf{16}, JFST0021 (2021).

\bibitem{Gomez2022}
M.~Gomez, S.~J. Grauer, J.~Ludwigsen, A.~M. Steinberg, S.~F. Son, S.~Roy, and
  T.~R. Meyer, \enquote{Megahertz-rate background-oriented schlieren tomography
  in post-detonation blasts,} Appl. Opt. \textbf{61}, 2444--2458 (2022).

\bibitem{Grauer2018}
S.~J. Grauer, A.~Unterberger, A.~Rittler, K.~J. Daun, A.~M. Kempf, and
  K.~Mohri, \enquote{Instantaneous {3D} flame imaging by background-oriented
  schlieren tomography,} Combust. Flame \textbf{196}, 284--299 (2018).

\bibitem{Liu2022}
L.~Liu, S.~Liu, H.~Yong, F.~Xiong, and T.~Yu, \enquote{Discontinuity computing
  with physics-informed neural network,} arXiv preprint p. 2206.03864 (2022).

\bibitem{Tokgoz2012}
S.~Tokgoz, R.~Geisler, L.~Van~Bokhoven, and B.~Wieneke, \enquote{Temperature
  and velocity measurements in a fluid layer using background-oriented
  schlieren and {PIV} methods,} Meas. Sci. Technol. \textbf{23}, 115302 (2012).

\bibitem{Dalziel2000}
S.~B. Dalziel, G.~O. Hughes, and B.~R. Sutherland, \enquote{Whole-field density
  measurements by {`synthetic schlieren'},} Exp. Fluids \textbf{28}, 322--335
  (2000).

\bibitem{Rajendran2020}
L.~K. Rajendran, J.~Zhang, S.~Bhattacharya, S.~P. Bane, and P.~P. Vlachos,
  \enquote{Uncertainty quantification in density estimation from
  background-oriented {Schlieren} measurements,} Meas. Sci. Technol.
  \textbf{31}, 054002 (2020).

\bibitem{Nicolas2016}
F.~Nicolas, V.~Todoroff, A.~Plyer, G.~Le~Besnerais, D.~Donjat, F.~Micheli,
  F.~Champagnat, P.~Cornic, and Y.~Le~Sant, \enquote{A direct approach for
  instantaneous {3D} density field reconstruction from background-oriented
  schlieren ({BOS}) measurements,} Exp. Fluids \textbf{57}, 1--21 (2016).

\bibitem{Grauer2020}
S.~J. Grauer and A.~M. Steinberg, \enquote{Fast and robust volumetric
  refractive index measurement by unified background-oriented schlieren
  tomography,} Exp. Fluids \textbf{61}, 1--17 (2020).

\bibitem{Castner2012}
R.~Castner, \enquote{Exhaust nozzle plume effects on sonic boom,} J. Aircr.
  \textbf{49}, 415--422 (2012).

\bibitem{Geerts2017}
J.~S. Geerts and K.~H. Yu, \enquote{Systematic application of
  background-oriented schlieren for isolator shock train visualization,} AIAA
  J. \textbf{55}, 1105--1117 (2017).

\bibitem{Rajendran2019}
L.~K. Rajendran, S.~P. Bane, and P.~P. Vlachos, \enquote{Dot tracking
  methodology for background-oriented schlieren ({BOS}),} Exp. Fluids
  \textbf{60}, 1--13 (2019).

\bibitem{Atcheson2009}
B.~Atcheson, W.~Heidrich, and I.~Ihrke, \enquote{An evaluation of optical flow
  algorithms for background oriented schlieren imaging,} Exp. Fluids
  \textbf{46}, 467--476 (2009).

\bibitem{Heineck2020}
J.~T. Heineck, D.~W. Banks, N.~T. Smith, E.~T. Schairer, P.~S. Bean, and
  T.~Robillos, \enquote{Background-oriented schlieren imaging of supersonic
  aircraft in flight,} AIAA J. \textbf{59}, 11--21 (2021).

\bibitem{Schmidt2021}
B.~E. Schmidt and M.~R. Woike, \enquote{Wavelet-based optical flow analysis for
  background-oriented schlieren image processing,} AIAA J. \textbf{59},
  3209--3216 (2021).

\bibitem{Raffel1998}
M.~Raffel, C.~E. Willert, J.~Kompenhans \emph{et~al.}, \emph{Particle Image
  Velocimetry: A Practical Guide}, vol.~2 (Springer, 1998).

\bibitem{Lucas1981}
B.~D. Lucas, T.~Kanade \emph{et~al.}, \enquote{An iterative image registration
  technique with an application to stereo vision,} in \enquote{DARPA Image
  Understanding Workshop,}  (1981), pp. 121--130.

\bibitem{Horn1981}
B.~K. Horn and B.~G. Schunck, \enquote{Determining optical flow,} Artif.
  Intell. \textbf{17}, 185--203 (1981).

\bibitem{Daun2016}
K.~J. Daun, S.~J. Grauer, and P.~J. Hadwin, \enquote{Chemical species
  tomography of turbulent flows: {Discrete} ill-posed and rank deficient
  problems and the use of prior information,} J. Quant. Spectrosc. Radiat.
  Transfer \textbf{172}, 58--74 (2016).

\bibitem{Kogelschatz1972}
U.~Kogelschatz and W.~Schneider, \enquote{Quantitative schlieren techniques
  applied to high current arc investigations,} Appl. Opt. \textbf{11},
  1822--1832 (1972).

\bibitem{Agrawal1999}
A.~K. Agrawal, B.~W. Albers, and D.~W. Griffin, \enquote{Abel inversion of
  deflectometric measurements in dynamic flows,} Appl. Opt. \textbf{38},
  3394--3398 (1999).

\bibitem{Daun2006}
K.~J. Daun, K.~A. Thomson, F.~Liu, and G.~J. Smallwood, \enquote{Deconvolution
  of axisymmetric flame properties using {Tikhonov} regularization,} Appl. Opt.
  \textbf{45}, 4638--4646 (2006).

\bibitem{Howard2016}
M.~Howard, M.~Fowler, A.~Luttman, S.~E. Mitchell, and M.~C. Hock,
  \enquote{Bayesian {Abel} inversion in quantitative {X}-ray radiography,} SIAM
  J. Sci. Comput. \textbf{38}, B396--B413 (2016).

\bibitem{Vauhkonen1998}
M.~Vauhkonen, D.~Vad{\'a}sz, P.~A. Karjalainen, E.~Somersalo, and J.~P. Kaipio,
  \enquote{Tikhonov regularization and prior information in electrical
  impedance tomography,} IEEE Trans. Med. Imaging \textbf{17}, 285--293 (1998).

\bibitem{Kolehmainen1998}
V.~Kolehmainen, E.~Somersalo, P.~Vauhkonen, M.~Vauhkonen, and J.~Kaipio,
  \enquote{A {Bayesian} approach and total variation priors in {3D} electrical
  impedance tomography,} in \enquote{Proceedings of the 20th Annual
  International Conference of the IEEE Engineering in Medicine and Biology
  Society,}  (IEEE, 1998), pp. 1028--1031.

\bibitem{Wei2021}
C.~Wei, K.~K. Schwarm, D.~I. Pineda, and R.~M. Spearrin, \enquote{Volumetric
  laser absorption imaging of temperature, {CO} and {CO$_2$} in laminar flames
  using {3D} masked {Tikhonov} regularization,} Combust. Flame \textbf{224},
  239--247 (2021).

\bibitem{Raissi2019}
M.~Raissi, P.~Perdikaris, and G.~E. Karniadakis, \enquote{Physics-informed
  neural networks: {A} deep learning framework for solving forward and inverse
  problems involving nonlinear partial differential equations,} J. Comp. Phys.
  \textbf{378}, 686--707 (2019).

\bibitem{Cai2021}
S.~Cai, Z.~Wang, F.~Fuest, Y.~J. Jeon, C.~Gray, and G.~E. Karniadakis,
  \enquote{Flow over an espresso cup: inferring {3-D} velocity and pressure
  fields from tomographic background oriented {Schlieren} via physics-informed
  neural networks,} J. Fluid Mech. \textbf{915} (2021).

\bibitem{Molnar2022}
J.~P. Molnar and S.~J. Grauer, \enquote{Flow field tomography with uncertainty
  quantification using a {Bayesian} physics-informed neural network,} Meas.
  Sci. Technol. \textbf{33}, 065305 (2022).

\bibitem{Mao2020}
Z.~Mao, A.~D. Jagtap, and G.~E. Karniadakis, \enquote{Physics-informed neural
  networks for high-speed flows,} Comput. Methods Appl. Mech. Eng.
  \textbf{360}, 112789 (2020).

\bibitem{Jagtap2022}
A.~D. Jagtap, Z.~Mao, N.~Adams, and G.~E. Karniadakis,
  \enquote{Physics-informed neural networks for inverse problems in supersonic
  flows,} J. Comput. Phys. \textbf{466}, 111402 (2022).

\bibitem{Gardiner1981}
W.~Gardiner~Jr, Y.~Hidaka, and T.~Tanzawa, \enquote{Refractivity of combustion
  gases,} Combust. Flame \textbf{40}, 213--219 (1981).

\bibitem{Born2013}
M.~Born and E.~Wolf, \emph{Principles of Optics: Electromagnetic Theory of
  Propagation, Interference and Diffraction of Light} (Elsevier, 2013).

\bibitem{Stam1996}
J.~Stam and E.~Langu{\'e}nou, \enquote{Ray tracing in non-constant media,} in
  \enquote{Eurographics Workshop on Rendering Techniques,}  (Springer, 1996),
  pp. 225--234.

\bibitem{Atcheson2008}
B.~Atcheson, I.~Ihrke, W.~Heidrich, A.~Tevs, D.~Bradley, M.~Magnor, and H.-P.
  Seidel, \enquote{Time-resolved {3D} capture of non-stationary gas flows,} ACM
  Trans. Graphics \textbf{27}, 1--9 (2008).

\bibitem{Goldhahn2007}
E.~Goldhahn and J.~Seume, \enquote{The background oriented schlieren technique:
  sensitivity, accuracy, resolution and application to a three-dimensional
  density field,} Exp. Fluids \textbf{43}, 241--249 (2007).

\bibitem{Sipkens2021}
T.~Sipkens, S.~Grauer, A.~Steinberg, S.~Rogak, and P.~Kirchen, \enquote{New
  transform to project axisymmetric deflection fields along arbitrary rays,}
  Meas. Sci. Technol. \textbf{33}, 035201 (2021).

\bibitem{Szeliski2010}
R.~Szeliski, \emph{Computer Vision: Algorithms and Applications} (Springer
  Science \& Business Media, 2010).

\bibitem{Davies2004}
E.~R. Davies, \emph{Machine Vision: Theory, Algorithms, Practicalities}
  (Elsevier, 2004).

\bibitem{Schmidt2019}
B.~Schmidt and J.~Sutton, \enquote{High-resolution velocimetry from tracer
  particle fields using a wavelet-based optical flow method,} Exp. Fluids
  \textbf{60}, 1--17 (2019).

\bibitem{Schmidt2020b}
B.~Schmidt and J.~Sutton, \enquote{Improvements in the accuracy of
  wavelet-based optical flow velocimetry ({wOFV}) using an efficient and
  physically based implementation of velocity regularization,} Exp. Fluids
  \textbf{61}, 1--17 (2020).

\bibitem{Corpetti2006}
T.~Corpetti, D.~Heitz, G.~Arroyo, E.~M{\'e}min, and A.~Santa-Cruz,
  \enquote{Fluid experimental flow estimation based on an optical-flow scheme,}
  Exp. Fluids \textbf{40}, 80--97 (2006).

\bibitem{Yuan2005}
J.~Yuan, P.~Ruhnau, E.~M{\'e}min, and C.~Schn{\"o}rr, \enquote{Discrete
  orthogonal decomposition and variational fluid flow estimation,} in
  \enquote{International Conference on Scale-Space Theories in Computer
  Vision,}  (Springer, 2005), pp. 267--278.

\bibitem{Kadri2013}
S.~Kadri-Harouna, P.~D{\'e}rian, P.~H{\'e}as, and E.~M{\'e}min,
  \enquote{Divergence-free wavelets and high order regularization,} Int. J.
  Comput. Vision \textbf{103}, 80--99 (2013).

\bibitem{Grauer2023}
S.~J. Grauer, K.~Mohri, T.~Yu, H.~Liu, and W.~Cai, \enquote{Volumetric emission
  tomography for combustion processes,} Prog. Energy Combust. Sci. \textbf{94},
  101024 (2023).

\bibitem{Ota2015}
M.~Ota, F.~Leopold, R.~Noda, and K.~Maeno, \enquote{Improvement in spatial
  resolution of background-oriented schlieren technique by introducing a
  telecentric optical system and its application to supersonic flow,} Exp.
  Fluids \textbf{56}, 1--10 (2015).

\bibitem{Hirose2019}
Y.~Hirose, K.~Ishikawa, Y.~Ishimoto, T.~Nagashima, M.~Ota, S.~Udagawa,
  T.~Inage, K.~Fujita, H.~Kiritani, K.~Fujita \emph{et~al.}, \enquote{The
  quantitative density measurement of unsteady flow around the projectile,} J.
  Flow Control Meas. Visualization \textbf{7}, 111 (2019).

\bibitem{Kolhe2009}
P.~S. Kolhe and A.~K. Agrawal, \enquote{Abel inversion of deflectometric data:
  comparison of accuracy and noise propagation of existing techniques,} Appl.
  Opt. \textbf{48}, 3894--3902 (2009).

\bibitem{Hayase2015}
T.~Hayase, \enquote{Numerical simulation of real-world flows,} Fluid Dyn. Res.
  \textbf{47}, 051201 (2015).

\bibitem{Cornick2009}
M.~Cornick, B.~Hunt, E.~Ott, H.~Kurtuldu, and M.~F. Schatz, \enquote{State and
  parameter estimation of spatiotemporally chaotic systems illustrated by an
  application to {Rayleigh--B{\'e}nard} convection,} Chaos Int. J. Nonlinear
  Sci. \textbf{19}, 013108 (2009).

\bibitem{Ali2022}
M.~Y.~B. Ali, O.~L{\'e}on, D.~Donjat, H.~B{\'e}zard, E.~Laroche, V.~Mons, and
  F.~Champagnat, \enquote{Data assimilation for aerothermal mean flow
  reconstruction using aero-optical observations: a synthetic investigation,}
  in \enquote{56 th 3AF International Conference on Applied Aerodynamics,}
  (2022), p.~11.

\bibitem{Saredi2021}
E.~Saredi, N.~T. Ramesh, A.~Sciacchitano, and F.~Scarano, \enquote{State
  observer data assimilation for {RANS} with time-averaged {3D-PIV} data,}
  Comput. Fluids \textbf{218}, 104827 (2021).

\bibitem{Mons2021}
V.~Mons, Y.~Du, and T.~A. Zaki, \enquote{Ensemble-variational assimilation of
  statistical data in large-eddy simulation,} Phys. Rev. Fluids \textbf{6},
  104607 (2021).

\bibitem{Wang2022b}
Q.~Wang, M.~Wang, and T.~A. Zaki, \enquote{What is observable from wall data in
  turbulent channel flow?} J. Fluid Mech. \textbf{941} (2022).

\bibitem{Vinnichenko2022}
N.~A. Vinnichenko, Y.~Y. Plaksina, A.~V. Pushtaev, and A.~V. Uvarov,
  \enquote{Obtaining velocity and pressure distributions in natural convection
  flows using experimental temperature fields,} Appl. Therm. Eng. p. 118962
  (2022).

\bibitem{Wang2022a}
S.~Wang, X.~Yu, and P.~Perdikaris, \enquote{When and why {PINNs} fail to train:
  {A} neural tangent kernel perspective,} J. Comp. Phys. \textbf{449}, 110768
  (2022).

\bibitem{Wang2021}
S.~Wang, Y.~Teng, and P.~Perdikaris, \enquote{Understanding and mitigating
  gradient flow pathologies in physics-informed neural networks,} SIAM J. Sci.
  Comput. \textbf{43}, A3055--A3081 (2021).

\bibitem{Jin2021}
X.~Jin, S.~Cai, H.~Li, and G.~E. Karniadakis, \enquote{{NSFnets}
  ({Navier--Stokes} flow nets): {Physics}-informed neural networks for the
  incompressible {Navier--Stokes} equations,} J. Comput. Phys. \textbf{426},
  109951 (2021).

\bibitem{Basir2022}
S.~Basir and I.~Senocak, \enquote{Physics and equality constrained artificial
  neural networks: Application to forward and inverse problems with
  multi-fidelity data fusion,} J. Comp. Phys. p. 111301 (2022).

\bibitem{Abadi2016}
M.~Abadi, P.~Barham, J.~Chen, Z.~Chen, A.~Davis, J.~Dean, M.~Devin,
  S.~Ghemawat, G.~Irving, M.~Isard \emph{et~al.}, \enquote{{TensorFlow}: a
  system for large-scale machine learning,} in \enquote{12th USENIX symposium
  on operating systems design and implementation (OSDI 16),}  (2016), pp.
  265--283.

\bibitem{Kingma2014}
D.~P. Kingma and J.~Ba, \enquote{Adam: {A} method for stochastic optimization,}
  arXiv preprint p. 1412.6980 (2014).

\bibitem{Cai2022}
S.~Cai, Z.~Mao, Z.~Wang, M.~Yin, and G.~E. Karniadakis,
  \enquote{Physics-informed neural networks ({PINNs}) for fluid mechanics: {A}
  review,} Acta Mech. Sin. pp. 1--12 (2022).

\bibitem{Casper2016}
K.~M. Casper, S.~J. Beresh, J.~F. Henfling, R.~W. Spillers, B.~O. Pruett, and
  S.~P. Schneider, \enquote{Hypersonic wind-tunnel measurements of
  boundary-layer transition on a slender cone,} AIAA J. \textbf{54}, 1250--1263
  (2016).

\bibitem{Walsh2000}
K.~T. Walsh, J.~Fielding, and M.~B. Long, \enquote{Effect of light-collection
  geometry on reconstruction errors in {Abel} inversions,} Opt. Lett.
  \textbf{25}, 457--459 (2000).

\bibitem{Sims1964}
J.~L. Sims, \emph{Tables for Supersonic Flow Around Right Circular Cones at
  Zero Angle of Attack}, vol. 3004 (Office of Scientific and Technical
  Information, National Aeronautics and Space Administration, 1964).

\bibitem{Cook1984}
R.~L. Cook, T.~Porter, and L.~Carpenter, \enquote{Distributed ray tracing,} in
  \enquote{Proceedings of the 11th annual conference on Computer graphics and
  interactive techniques,}  (1984), pp. 137--145.

\bibitem{Sharma1982}
A.~Sharma, D.~V. Kumar, and A.~K. Ghatak, \enquote{Tracing rays through
  graded-index media: a new method,} Appl. Opt. \textbf{21}, 984--987 (1982).

\bibitem{Foi2008}
A.~Foi, M.~Trimeche, V.~Katkovnik, and K.~Egiazarian, \enquote{Practical
  {Poissonian-Gaussian} noise modeling and fitting for single-image raw-data,}
  IEEE Trans. Image Process. \textbf{17}, 1737--1754 (2008).

\bibitem{Economon2016}
T.~D. Economon, F.~Palacios, S.~R. Copeland, T.~W. Lukaczyk, and J.~J. Alonso,
  \enquote{{SU2}: {An} open-source suite for multiphysics simulation and
  design,} AIAA J. \textbf{54}, 828--846 (2016).

\bibitem{Anderson1990}
J.~D. Anderson, \emph{Modern Compressible Flow: With Historical Perspective},
  vol.~12 (McGraw-Hill New York, 1990).

\bibitem{Haghighat2022}
E.~Haghighat, D.~Amini, and R.~Juanes, \enquote{Physics-informed neural network
  simulation of multiphase poroelasticity using stress-split sequential
  training,} Comput. Methods Appl. Mech. Eng. \textbf{397}, 115141 (2022).

\bibitem{Cormack1982}
A.~M. Cormack, \enquote{Computed tomography: some history and recent
  developments,} in \enquote{Proceedings of Symposia in Applied Mathematics,}
  (American Mathematical Society, 1982), pp. 35--42.

\bibitem{Neumann1950}
J.~von Neumann and R.~D. Richtmyer, \enquote{A method for the numerical
  calculation of hydrodynamic shocks,} J. Appl. Phys. \textbf{21}, 232--237
  (1950).

\bibitem{Lax1959}
P.~Lax, \enquote{Systems of conservation laws,} Tech. rep., Los Alamos National
  Lab (1959).

\bibitem{Fuks2020}
O.~Fuks and H.~A. Tchelepi, \enquote{Limitations of physics informed machine
  learning for nonlinear two-phase transport in porous media,} J. Mach. Learn.
  Model. Comput. \textbf{1} (2020).

\bibitem{Coutinho2022}
E.~J.~R. Coutinho, M.~Dall'Aqua, L.~McClenny, M.~Zhong, U.~Braga-Neto, and
  E.~Gildin, \enquote{Physics-informed neural networks with adaptive localized
  artificial viscosity,} arXiv preprint p. 2203.08802 (2022).

\bibitem{Patel2022}
R.~G. Patel, I.~Manickam, N.~A. Trask, M.~A. Wood, M.~Lee, I.~Tomas, and E.~C.
  Cyr, \enquote{Thermodynamically consistent physics-informed neural networks
  for hyperbolic systems,} J. Comp. Phys. \textbf{449}, 110754 (2022).

\bibitem{Lellis2010}
C.~De~Lellis and L.~Sz{\'e}kelyhidi, \enquote{On admissibility criteria for
  weak solutions of the {Euler} equations,} Arch. Ration. Mech. Anal.
  \textbf{195}, 225--260 (2010).

\bibitem{Godlewski2013}
E.~Godlewski and P.-A. Raviart, \emph{Numerical Approximation of Hyperbolic
  Systems of Conservation Laws}, vol. 118 (Springer Science \& Business Media,
  2013).

\bibitem{Roe1981}
P.~L. Roe, \enquote{Approximate {Riemann} solvers, parameter vectors, and
  difference schemes,} J. Comput. Phys. \textbf{43}, 357--372 (1981).

\bibitem{Venkatakrishnan1995}
V.~Venkatakrishnan, \enquote{Convergence to steady state solutions of the
  {Euler} equations on unstructured grids with limiters,} J. Comput. Phys.
  \textbf{118}, 120--130 (1995).

\end{thebibliography}

\end{document}